\documentclass[12pt,english]{article}
\usepackage[T1]{fontenc}
\usepackage[latin9]{inputenc}
\usepackage{amsmath}
\usepackage{amssymb}
\usepackage{graphicx}
\usepackage{color}

\makeatletter
\usepackage{jheppub}
\def\@fpheader{\relax}
\allowdisplaybreaks[1]

\makeatother

\usepackage{babel}
\begin{document}

\subheader{}

\title{Black hole singularity resolution via the modified Raychaudhuri equation in loop quantum gravity}

\author[a]{Keagan Blanchette,}
\author[b]{Saurya Das,}
\author[a]{Samantha Hergott,}
\author[a]{Saeed Rastgoo,}
\affiliation[a]{Department of Physics and Astronomy, York University~\\  4700 Keele Street, Toronto, Ontario M3J 1P3 Canada} 
\affiliation[b]{Theoretical Physics Group and Quantum Alberta, Department of Physics and Astronomy, University of Lethbridge, 4401 University Drive, Lethhbridge, Alberta T1K 3M4, Canada}
\emailAdd{kblanch@yorku.ca}
\emailAdd{saurya.das@uleth.ca}
\emailAdd{sherrgs@yorku.ca}
\emailAdd{srastgoo@yorku.ca}

\abstract{We derive loop quantum gravity corrections to the Raychaudhuri equation
in the interior of a Schwarzschild black hole and near the classical singularity.
We show that the resulting effective equation implies defocusing of geodesics due to the appearance of repulsive terms.
This prevents the formation of conjugate points,   
renders the singularity theorems inapplicable,
and leads to the resolution of the singularity for this spacetime.}

\maketitle

\section{Introduction}

It is well known that General Relativity (GR) predicts that all reasonable spacetimes are singular, and therefore its own demise. 
While a similar situation in electrodynamics was resolved in quantum electrodynamics, quantum gravity has not been completely formulated yet. One of the primary challenges of candidate theories such as 
string theory and loop quantum gravity (LQG) is to  find a way of resolving the singularities.

Singularities in GR are defined differently compared to other field theories. While curvature scalars (such as the Kretschmann scalar) approaching infinity (similar to the electric field diverging at the seat of a charge)
is a strong indication of singularities, it is neither a necessary nor a sufficient condition for singularities in GR. 
The necessary and sufficient condition for a singular spacetime is the existence of a set of geodesics which begin and/or end at a finite proper time. Such geodesics are deemed incomplete. Furthermore, the celebrated Hawking-Penrose singularity theorems prove beyond doubt that under normal assumptions, all spacetime solutions of GR will have incomplete geodesics, and will therefore be  singular
\cite{Penrose:1964wq,Hawking:1969sw,Raychaudhuri:1953yv}.

It may be mentioned that the proof of the singularity theorems crucially depend on the fact that there exist congruences or a collection of nearby geodesics, such that they focus to the conjugate points in the past as well as in the future at finite proper times. This implies that the geodesics are no longer maximal curves in a pseudo-Riemannian manifold, contrary to their very definition as solutions of the geodesic equation. The only resolution of this apparent contradiction is to conclude that such geodesics are incomplete. The existence of conjugate points is a straightforward prediction of the Raychaudhuri equation \cite{Raychaudhuri:1953yv}. 

In view of the above, in this article, we examine the issue of singularity resolution via the LQG modified Raychaudhuri equation, and 
in particular for the Schwarzschild solution in GR. Since the classical singularity is at the origin of the above black hole metric, $r=0$, we focus on the region inside the classical horizon at $r=2GM$, where $M$ is the mass of the black hole and $G$ is Newton's constant (we work in $c=1=\hbar$ units). By choosing the appropriate regularized tetrads and holonomies, which are the conjugate variables in LQG, computing the corresponding expansion of geodesics and substituting in the Raychaudhuri equation, we show that they include effective repulsive terms which prevents the formation of conjugate points. This implies that the classical singularity theorems are rendered invalid and the singularity is resolved, at least for the spacetime under consideration. While our results strictly pertain to the static Schwarzschild spacetime, the robustness of our results indicate that the resolution will continue to hold for more complicated spacetimes, including those with little or no symmetries, and when quantum corrections from other sources are taken into account (e.g., in Refs. \cite{Das:2013oda,Burger:2018hpz,Das:2019vnx}).

As is well known, LQG \cite{Thiemann:2007pyv} is one of the
main nonperturbative approaches to the quantization of gravity. Within
LQG, there have been numerous studies of both the interior and the
full spacetime of black holes in four and lower dimensions \cite{Bojowald:2004af,Ashtekar:2005qt,Bojowald:2005cb,Bohmer:2007wi,Boehmer:2008fz,Corichi:2015xia,BenAchour:2017ivq,Ashtekar:2018cay,BenAchour:2018khr,Barrau:2018rts,Aruga:2019dwq,BenAchour:2020gon,Bodendorfer:2019cyv,Bodendorfer:2019nvy,Bojowald:2008bt,Bojowald:2008ja,Bojowald:2016itl,Bojowald:2016vlj,Bojowald:2018xxu,Brahma:2014gca,Campiglia:2007pb,Chiou:2008nm,Corichi:2015vsa,Cortez:2017alh,Gambini:2008dy,Gambini:2009ie,Gambini:2011mw,Gambini:2013ooa,Gambini:2020nsf,Husain:2004yz,Husain:2006cx,Kelly:2020lec,Kelly:2020uwj,Kreienbuehl:2010vc,Modesto:2005zm,Modesto:2009ve,Olmedo:2017lvt,Thiemann:1992jj,Zhang:2020qxw,Ziprick:2016ogy,Campiglia:2007pr,Gambini:2009vp,Rastgoo:2013isa,Corichi:2016nkp,Morales-Tecotl:2018ugi,BenAchour:2020bdt,BenAchour:2020mgu}.
These attempts were originally inspired by loop quantum cosmology
(LQC), more precisely a certain quantization of the isotropic Friedmann-Lemaitre-Robertson-Walker (FLRW) model
\cite{Ashtekar:2006rx,Ashtekar:2006uz} which uses a certain type
of quantization of the phase space called polymer quantization \cite{Ashtekar:2002sn,Corichi:2007tf,Morales-Tecotl:2016ijb,Tecotl:2015cya, Flores-Gonzalez:2013zuk}.
This quantiztion introduces a parameter into the theory called the
polymer scale that sets the minimal scale of the model. Close to this scale quantum effects become important. The approach in which such a parameter
is taken to be constant is called the $\mu_{0}$ scheme (which in
this paper we refer to as the $\mathring{\mu}$ scheme), while approaches
where it depends on the phase space variables are denoted by $\bar{\mu}$
schemes. These various approaches were introduced to deal with some
important issues resulting from quantization, namely, to have the
correct classical limit (particularly in LQC), to avoid large quantum
corrections near the horizon, and to have final physical results that
are independent of auxiliary or fiducial parameters. Other approaches to this model in LQG such as Refs. \cite{Alesci:2018loi,Alesci:2019pbs,Alesci:2020zfi} provide a derivation of a Schwarzschild black hole modified dynamics for the interior and the exterior regions, not relying on minisuperspace models. Starting from the full LQG theory, this model performs the symmetry reduction at the quantum level. This has led to several differences in the effective dynamics with respect to previous polymer quantization-inspired models, one of which is the absence of the formation of a white hole in the extended spacetime region replacing the classical singularity. All of these past studies in LQG and some other approaches (see, e.g., Refs. \cite{Saini:2014qpa,Greenwood:2008ht,Wang:2009ay}) point to the resolution of the singularity at the effective level. 

In this paper, we consider the interior of the Schwarzschild black
hole expressed in terms of connection variables and follow the same
polymer quantization as previous works based on minisuperspace models but study the behavior of
modified effective geodesics in the interior of the black hole using
the modified Raychaudhuri equation. We will consider both the $\mathring{\mu}$
scheme and two of the most common cases in $\bar{\mu}$ schemes.

This paper is organized as follows. In Sec. \ref{sec:Schw-int}, we
review the classical interior of the Schwarzschild black hole. In
Sec. \ref{sec:Classical-Dynamics}, we remind the reader of the classical
dynamics of the interior, derive the corresponding Raychaudhuri
equation, and show that this leads to the expected presence of a singularity
at the center of the black hole. In Sec. \ref{sec:Effective-Dynamics},
we present the effective dynamics of the interior after polymer quantization
in a general setting. We then go on  to derive the effective Raychaudhuri
equation of the $\mathring{\mu}$ scheme and two of the most common
$\bar{\mu}$ schemes in Secs. \ref{subsec:mu0-scheme}, \ref{subsec:mubar-1-scheme},
and \ref{subsec:mubar-2-scheme}, respectively, showing how the modified
behavior of geodesics shows the resolution of singularity in each of
these cases.

\section{Interior of the Schwarzschild black hole\label{sec:Schw-int}}

The celebrated metric of the exterior of a Schwarzschild black hole
of mass $M$ is given by

\begin{equation}
ds^{2}=-\left(1-\frac{2GM}{r}\right)dt^{2}+\left(1-\frac{2GM}{r}\right)^{-1}dr^{2}+r^{2}\left(d\theta^{2}+\sin^{2}\theta d\phi^{2}\right),
\end{equation}
where and $r\in(0,\infty)$ is the radial coordinate distance and
the radius of the 2-spheres in Schwarzschild coordinates $\left(t,r,\theta,\phi\right)$.
It is well known that for such a black hole, the timelike and spacelike
curves switch their causal nature upon crossing the event horizon
located at $R_{s}=2GM$. 
Thus,
the interior metric can be written as 
\begin{equation}
ds^{2}=-\left(\frac{2GM}{t}-1\right)^{-1}dt^{2}+\left(\frac{2GM}{t}-1\right)dr^{2}+t^{2}\left(d\theta^{2}+\sin^{2}\theta d\phi^{2}\right).\label{eq:sch-inter}
\end{equation}
Here and throughout the paper, $t$ is the Schwarzschild time with
the range $t\in(0,2GM)$. This metric is a special case of a Kantowski-Sachs
cosmological spacetime that is given by the metric \cite{Collins:1977fg} 
\begin{align}
ds_{KS}^{2}= & -N(T)^{2}dT^{2}+g_{xx}(T)dx^{2}+g_{\theta\theta}(T)d\theta^{2}+g_{\phi\phi}(T)d\phi^{2}\nonumber \\
= & -d\tau^{2}+g_{xx}(\tau)dx^{2}+g_{\Omega\Omega}(\tau)d\Omega^{2}.\label{eq:K-S-gener}
\end{align}
Note that $x$ here is not necessarily the radius $r$ of the 2-spheres
with area $A=4\pi r^{2}$, 
where $N(T)$ is the lapse function corresponding to a generic time $T$. It is seen that the metric (\ref{eq:sch-inter}) and (\ref{eq:K-S-gener})
are related by a transformation, 
\begin{equation}
d\tau^{2}=N(T)^{2}dT^{2}=\left(\frac{2GM}{t}-1\right)^{-1}dt^{2}.
\end{equation}
The metric (\ref{eq:K-S-gener}) represents a spacetime with spatial
homogeneous but anisotropic foliations. A quick way to see this is
that $g_{xx}(\tau)$ and $g_{\Omega\Omega}(\tau)$ can be considered
as two distinct scale factors that affect the radial and angular parts
of the metric separately. As is evident from (\ref{eq:K-S-gener}),
such a system is a minisuperspace model due to incorporating a finite
number of configuration variables. Furthermore, it can be seen that
the spatial hypersurfaces have topology $\mathbb{R}\times\mathbb{S}^{2}$,
and the spatial symmetry group is the Kantowski-Sachs isometry group
$\mathbb{R}\times SO(3)$. Due to this topology with a noncompact
direction, $x\in\mathbb{R}$ in space, the symplectic form $\int_{\mathbb{R}\times\mathbb{S}^{2}}\text{d}^{3}x\,\text{d}q\wedge\text{d}p$
diverges. Therefore, one needs to choose a finite fiducial volume
over which this integral is calculated \cite{Ashtekar:2005qt}. This
is a common practice in the study of homogeneous minisuperspace models.
Here one introduces an auxiliary length $L_{0}$ to restrict the noncompact
direction to an interval $x\in\mathcal{I}=[0,L_{0}]$. The volume
of the fiducial cylindrical cell in this case is $V_{0}=a_{0}L_{0}$,
where $a_{0}$ is the area of the 2-sphere $\mathbb{S}^{2}$ in $\mathcal{I}\times\mathbb{S}^{2}$. 

In order to obtain the Hamiltonian of this system in connection variables,
one first considers the full Hamiltonian of gravity written in terms
of (the curvature) of the $su(2)$ Ashtekar-Barbero connection $A_{a}^{i}$
and its conjugate momentum, the densitized triad $\tilde{E}_{a}^{i}$.
Using the Kantowski-Sachs symmetry, these variables can be written
as \cite{Ashtekar:2005qt}
\begin{align}
A_{a}^{i}\tau_{i}dx^{a}= & \frac{c}{L_{0}}\tau_{3}dx+b\tau_{2}d\theta-b\tau_{1}\sin\theta d\phi+\tau_{3}\cos\theta d\phi,\label{eq:A-AB}\\
\tilde{E}_{i}^{a}\tau_{i}\partial_{a}= & p_{c}\tau_{3}\sin\theta\partial_{x}+\frac{p_{b}}{L_{0}}\tau_{2}\sin\theta\partial_{\theta}-\frac{p_{b}}{L_{0}}\tau_{1}\partial_{\phi},\label{eq:E-AB}
\end{align}
where $b$, $c$, $p_{b}$, and $p_{c}$ are functions that only depend
on time and $\tau_{i}=-i\sigma_{i}/2$ are a $su(2)$ basis satisfying
$\left[\tau_{i},\tau_{j}\right]=\epsilon_{ij}{}^{k}\tau_{k}$, with
$\sigma_{i}$ being the Pauli matrices. Substituting these into the
full Hamiltonian of gravity written in Ashtekar connection variables,
one obtains the symmetry reduced Hamiltonian constraint adapted to
this model as \cite{Ashtekar:2005qt}
\begin{equation}
H=-\frac{N}{2G\gamma^{2}}\left[2bc\sqrt{p_{c}}+\left(b^{2}+\gamma^{2}\right)\frac{p_{b}}{\sqrt{p_{c}}}\right],
\end{equation}
while the diffeomorphism constraint vanishes identically due to the
homogenous nature of the model. Here, $\gamma$ is the Barbero-Immirzi
parameter \cite{Thiemann:2007pyv}, and $p_{c}\geq0$. $\gamma$ is
the term whose inverse couples the first order Palatini action to
a topological Nieh-Yan term yielding the Holst action. Hence it does
not affect the equations of motion and classically has no effect on
the system. However, after quantization, this parameter sets the size
of the quantum of area in Planck units.

Using symmetry of the model, the full symplectic form
\begin{equation}
\boldsymbol{\Omega}=\frac{1}{8\pi G\gamma}\int_{\mathcal{I}\times\mathbb{S}^{2}}d^{3}xdA_{a}^{i}(\mathbf{x})\wedge d\tilde{E}_{i}^{a}(\mathbf{y})
\end{equation}
reduces to \cite{Ashtekar:2005qt}
\begin{equation}
\boldsymbol{\Omega}=\frac{1}{2G\gamma}\left(dc\wedge dp_{c}+2db\wedge dp_{b}\right),
\end{equation}
 and consequently the Poisson brackets
\begin{equation}
\left\{ A_{a}^{i}(\mathbf{x}),\tilde{E}_{j}^{b}(\mathbf{y})\right\} =8\pi G\gamma\delta_{j}^{i}\delta_{a}^{b}\delta^{3}\left(\mathbf{x}-\mathbf{y}\right)
\end{equation}
 reduce to
\begin{equation}
\{c,p_{c}\}=2G\gamma,\quad\quad\{b,p_{b}\}=G\gamma.\label{eq:classic-PBs-bc}
\end{equation}
Furthermore, by substituting (\ref{eq:A-AB}) and (\ref{eq:E-AB}),
and the components of the inverse of the metric (\ref{eq:K-S-gener}),
into the relation between the inverse triad and the spatial metric
$q_{ab}$,
\begin{equation}
qq^{ab}=\delta^{ij}\tilde{E}_{i}^{a}\tilde{E}_{j}^{b},
\end{equation}
one obtains for the generic metric (\ref{eq:K-S-gener}) adapted to
(\ref{eq:A-AB}) and (\ref{eq:E-AB})
\begin{align}
g_{xx}\left(T\right)= & \frac{p_{b}\left(T\right)^{2}}{L_{0}^{2}p_{c}\left(T\right)},\label{eq:grrT}\\
g_{\theta\theta}\left(T\right)= & \frac{g_{\phi\phi}\left(T\right)}{\sin^{2}\left(\theta\right)}=g_{\Omega\Omega}\left(T\right)=p_{c}\left(T\right).\label{eq:gththT}
\end{align}

Note that the lapse $N(T)$ is not determined and can be chosen as
suited for a specific situation. Hence, the adapted metric using (\ref{eq:grrT})
and (\ref{eq:gththT}) becomes
\begin{equation}
ds^{2}=-N(T)^{2}dT^{2}+\frac{p_{b}^{2}}{L_{0}^{2}\,p_{c}}dx^{2}+p_{c}(d\theta^{2}+\sin^{2}\theta d\phi^{2}).
\end{equation}
Comparing this metric written in Schwarzschild coordinates and lapse
$N(t)$ with the standard Schwarzschild interior metric but with
rescaled $r\to lx,$
\begin{equation}
ds^{2}=-\left(\frac{2GM}{t}-1\right)^{-1}dt^{2}+l^{2}\left(\frac{2GM}{t}-1\right)dx^{2}+t^{2}\left(d\theta^{2}+\sin^{2}\theta d\phi^{2}\right),
\end{equation}
we see that 
\begin{align}
N\left(t\right)= & \left(\frac{2GM}{t}-1\right)^{-\frac{1}{2}},\label{eq:Sch-corresp-1}\\
g_{xx}\left(t\right)= & \frac{p_{b}\left(t\right)^{2}}{L_{0}^{2}p_{c}\left(t\right)}=l^{2}\left(\frac{2GM}{t}-1\right),\label{eq:Sch-corresp-2}\\
g_{\theta\theta}\left(T\right)= & \frac{g_{\phi\phi}\left(T\right)}{\sin^{2}\left(\theta\right)}=g_{\Omega\Omega}\left(T\right)=p_{c}\left(t\right)=t^{2}.\label{eq:Sch-corresp-3}
\end{align}
Hereafter we use $l=1$. This shows that 
\begin{align}
p_{b}= & 0, & p_{c}= & 4G^{2}M^{2}, &  & \textrm{on the horizon\,}t=2GM,\label{eq:t-horiz}\\
p_{b}\to & 0, & p_{c}\to & 0, &  & \textrm{at the singularity\,}t\to0.\label{eq:t-singular}
\end{align}
Also, note that in the fiducial volume, we can consider three surfaces
$S_{x,\theta},\,S_{x,\phi}$, and $S_{\theta,\phi}$, respectively, bounded
by $\mathcal{I}$ and a great circle along a longitude of $V_{0}$,
$\mathcal{I}$ and the equator of $V_{0}$, and the equator and a
longitude with areas \cite{Ashtekar:2005qt}
\begin{align}
A_{x,\theta}=A_{x,\phi}= & 2\pi L_{0}\sqrt{g_{xx}g_{\Omega\Omega}}=2\pi p_{b},\label{eq:area-x}\\
A_{\theta,\phi}= & \pi g_{\Omega\Omega}=\pi p_{c},\label{eq:area-angl}
\end{align}
with the volume of the fiducial region $\mathcal{I}\times\mathbb{S}^{2}$
given by \cite{Ashtekar:2005qt}
\begin{equation}
V=\int\mathrm{d}^{3}x\sqrt{|\det\tilde{E}|}=4\pi L\sqrt{g_{xx}}g_{\Omega\Omega}=4\pi p_{b}\sqrt{p_{c}},
\end{equation}
where $\sqrt{\det|\tilde{E}|}=\sqrt{q}$ with $q$ being the determinant
of the spatial metric.

\section{Classical dynamics\label{sec:Classical-Dynamics}}

\subsection{Classical Hamiltonian and equations of motion}

We are interested in the classical dynamics of the interior of the
Schwarzschild black hole in Ashtekar-Barbero connection formulation.
As usual in gravity, the classical Hamiltonian is the sum of constraints
that generate spacetime diffeomorphisms and internal or Gauss (in
our case $su(2)$) symmetry. The full version of the classical Hamiltonian
constraint in this formulation is \cite{Thiemann:2007pyv}
\begin{equation}
H_{\textrm{full}}=\frac{1}{8\pi G}\int d^{3}x\frac{N}{\sqrt{\det|\tilde{E}|}}\left\{ \epsilon_{i}^{jk}F_{ab}^{i}\tilde{E}_{j}^{a}\tilde{E}_{k}^{b}-2\left(1+\gamma^{2}\right)K_{[a}{}^{i}K_{b]}^{j}\tilde{E}_{i}^{a}\tilde{E}_{j}^{b}\right\}, \label{eq:Full-H-gr-class}
\end{equation}
where $K_{a}^{i}$ is the extrinsic curvature of foliations and $\epsilon_{ijk}$
is the totally antisymmetric Levi-Civita symbol. Also, $F=dA+A\wedge A$
is the curvature of the Ashtekar-Barbero connection. The symmetry
reduced Hamiltonian corresponding to the above full Hamiltonian is
derived by substituting (\ref{eq:A-AB}) and (\ref{eq:E-AB}) in (\ref{eq:Full-H-gr-class}).
In this way, one obtains \cite{Ashtekar:2005qt,Chiou:2008nm,Corichi:2015xia,Bohmer:2007wi,Morales-Tecotl:2018ugi}
\begin{equation}
H=-\frac{N}{2G\gamma^{2}}\left[\left(b^{2}+\gamma^{2}\right)\frac{p_{b}}{\sqrt{p_{c}}}+2bc\sqrt{p_{c}}\right].\label{eq:H-const-class}
\end{equation}
Given the homogeneous nature of the model, the diffeomorphism constraint
is trivially satisfied, and after imposing the Gauss constraint, one
is left only with the classical Hamiltonian constraint (\ref{eq:H-const-class}).
In order to facilitate the derivation of the solutions to the equations
of motion, we choose a gauge where the lapse function is 
\begin{equation}
N\left(T\right)=\frac{\gamma\sqrt{p_{c}\left(T\right)}}{b\left(T\right)},\label{eq:lapsNT}
\end{equation}
for which the Hamiltonian constraint becomes
\begin{equation}
H=-\frac{1}{2G\gamma}\left[\left(b^{2}+\gamma^{2}\right)\frac{p_{b}}{b}+2cp_{c}\right].\label{eq:H-const-cls-gauged}
\end{equation}
The advantage of this lapse function is that the equations of motion
of $c,\,p_{c}$ decouple from those of $b,\,p_{b}$, 
\begin{align}
\frac{db}{dT}= & \left\{ b,H\right\} =-\frac{1}{2}\left(b+\frac{\gamma^{2}}{b}\right),\label{eq:EoM-cls-b}\\
\frac{dp_{b}}{dT}= & \left\{ p_{b},H\right\} =\frac{p_{b}}{2}\left(1-\frac{\gamma^{2}}{b^{2}}\right),\label{eq:EoM-cls-pb}\\
\frac{dc}{dT}= & \left\{ c,H\right\} =-2c,\label{eq:EoM-cls-c}\\
\frac{dp_{c}}{dT}= & \left\{ p_{c},H\right\} =2p_{c}.\label{eq:EoM-cls-pc}
\end{align}
These equations are also to be supplemented with the on-shell condition
of the vanishing of the Hamiltonian constraint (\ref{eq:H-const-cls-gauged})
on the constraint surface
\begin{equation}
\left(b^{2}+\gamma^{2}\right)\frac{p_{b}}{b}+2cp_{c}\approx0,\label{eq:H-weak-vanish}
\end{equation}
where $\approx$ stands for weak equality, i.e., on the constraint
surface. 

It is clear from (\ref{eq:Sch-corresp-3}) that $p_{c}$ is the square
of the radius of the infalling 2-spheres. In order to better understand
the role of $b,\,c$ we use the relation of the proper time $\tau$
and a generic time $T$ for the metric (\ref{eq:K-S-gener}),
\begin{equation}
d\tau^{2}=-N(T)^{2}dT^{2},
\end{equation}
and the form of the lapse function (\ref{eq:lapsNT}), to rewrite
Eqs. (\ref{eq:EoM-cls-pc}) as
\begin{equation}
b=\frac{\gamma}{2}\frac{1}{\sqrt{p_{c}}}\frac{dp_{c}}{d\tau}=\gamma\frac{d}{d\tau}\sqrt{g_{\Omega\Omega}}=\frac{\gamma}{\sqrt{\pi}}\frac{d}{d\tau}\sqrt{A_{\theta,\phi}},
\end{equation}
where the last two terms on the right-hand side were derived using
(\ref{eq:area-angl}). Hence, classically, $b$ is proportional to
the rate of change of the square root of the physical area of $\mathbb{S}^{2}$. 

To interpret the role of $c$, we use the same method for (\ref{eq:EoM-cls-pc}),
and by using (\ref{eq:H-weak-vanish}) to replace $\gamma\frac{p_{b}}{b}=-\frac{bp_{b}}{\gamma}-\frac{2cp_{c}}{\gamma}$
in the resultant expression and then using (\ref{eq:area-x}) and
(\ref{eq:area-angl}), we find out
\begin{equation}
c=\gamma\frac{d}{d\tau}\left(\frac{p_{b}}{\sqrt{p_{c}}}\right)=\gamma\frac{d}{d\tau}\left(L_{0}\sqrt{g_{xx}}\right).
\end{equation}
Hence, classically, $c$ is proportional to the rate of change of the
physical length of $\mathcal{I}$.

\begin{figure}
\begin{centering}
\includegraphics[scale=0.7]{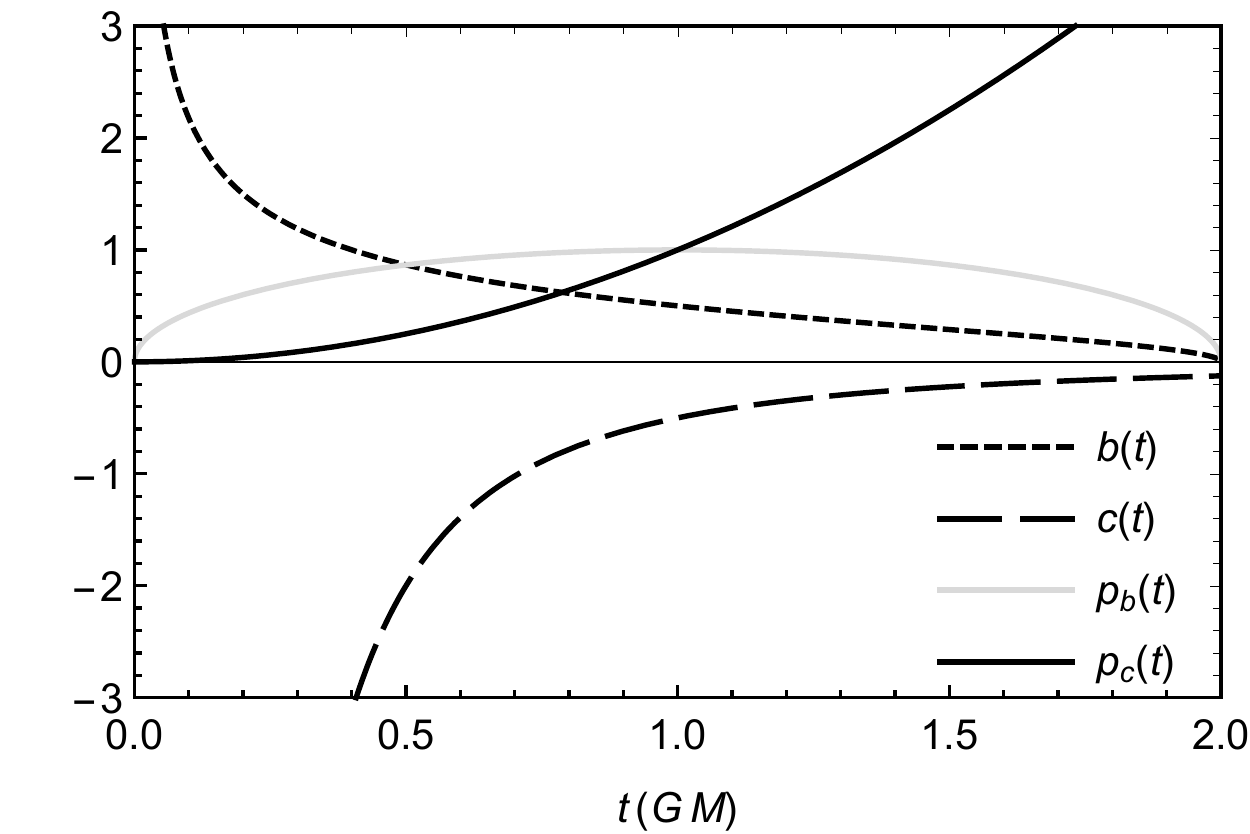}
\par\end{centering}
\caption{The behavior of canonical variables as a function of Schwarzschild
time $t$. We have chosen the positive sign for $b$ and negative
sign for $c$. The figure is plotted using $\gamma=0.5,\,M=1,\,G=1$,
and $L_{0}=1$. \label{fig:class-var-behv}}

\end{figure}

The solution to the classical equations of motion (\ref{eq:EoM-cls-b})--(\ref{eq:EoM-cls-pc})
are be found to be
\begin{align}
b\left(T\right)= & \pm\sqrt{e^{2C_{1}}e^{-T}-\gamma^{2}},\\
p_{b}\left(T\right)= & C_{2}e^{\frac{T}{2}}\sqrt{e^{2C_{1}}-\gamma^{2}e^{T}},\\
c\left(T\right)= & C_{3}e^{-2T},\\
p_{c}\left(T\right)= & C_{4}e^{2T}.
\end{align}
Since we know from (\ref{eq:Sch-corresp-3}) that in Schwarzschild
coordinates $p_{c}\left(t\right)=t^{2}$, and considering the fourth
equation above, we see that a transformation $T=\ln\left(t\right)$
can lead to such a solution for $p_{c}$. Under such a transformation,
the above equations become
\begin{align}
b\left(t\right)= & \pm\sqrt{\frac{e^{2C_{1}}}{t}-\gamma^{2}},\\
p_{b}\left(t\right)= & C_{2}t\sqrt{\frac{e^{2C_{1}}}{t}-\gamma^{2}},\label{eq:EoM-pbt-C2}\\
c\left(t\right)= & \frac{C_{3}}{t^{2}},\\
p_{c}\left(t\right)= & C_{4}t^{2}.
\end{align}
Considering (\ref{eq:Sch-corresp-3}), we see that
\begin{equation}
C_{4}=1.
\end{equation}
Also, from (\ref{eq:t-horiz}), we can deduce
\begin{equation}
0=p_{b}\left(2GM\right)=2GMC_{2}\sqrt{\frac{e^{2C_{1}}}{2GM}-\gamma^{2}},
\end{equation}
which yields
\begin{equation}
C_{1}=\frac{1}{2}\ln\left(2GM\gamma^{2}\right).\label{eq:C1-EoM}
\end{equation}
Next, from (\ref{eq:Sch-corresp-2}), we see
\begin{equation}
p_{b}\left(t\right)^{2}=l^{2}\left(\frac{2GM}{t}-1\right)L_{0}^{2}t^{2},
\end{equation}
which if compared with (\ref{eq:EoM-pbt-C2}) and using (\ref{eq:C1-EoM})
yields
\begin{equation}
C_{2}=\frac{lL_{0}}{\gamma}.
\end{equation}
Finally, using (\ref{eq:H-weak-vanish}), we get 
\begin{equation}
C_{3}=\mp\gamma GMlL_{0}.
\end{equation}
Hence, the equations of motion in Schwarzschild written in $t$ become
\begin{align}
b\left(t\right)= & \pm\gamma\sqrt{\frac{2GM}{t}-1},\label{eq:sol-cls-b}\\
p_{b}\left(t\right)= & lL_{0}t\sqrt{\frac{2GM}{t}-1},\label{eq:sol-cls-pb}\\
c\left(t\right)= & \mp\frac{\gamma GMlL_{0}}{t^{2}},\label{eq:sol-cls-c}\\
p_{c}\left(t\right)= & t^{2}.\label{eq:sol-cls-pc}
\end{align}
The behavior of these solutions as a function of $t$ is depicted
in Fig. \ref{fig:class-var-behv}. From these equations or the plot,
one can see that $p_{c}\to0$ as $t\to0$. This means that at the classical
singularity, the Riemann invariants such as the Kretschmann
scalar
\begin{equation}
K=R_{abcd}R^{abcd}\propto\frac{1}{p_{c}^{3}},
\end{equation}
all diverge, signaling the presence of a physical singularity there
as expected.

We can also see from Fig. \ref{fig:class-var-behv} that $b$, the
rate of change of the square root of the physical area of $\mathbb{S}^{2}$,
as well as $c$, the rate of change of the physical length of $\mathcal{I}$,
diverge at the classical singularity.

\subsection{Classical Raychaudhuri equation}

The celebrated Raychaudhuri equation \cite{Raychaudhuri:1953yv}
\begin{equation}
\frac{d\theta}{d\tau}=-\frac{1}{3}\theta^{2}-\sigma_{ab}\sigma^{ab}+\omega_{ab}\omega^{ab}-R_{ab}U^{a}U^{b}
\end{equation}
describes the behavior of geodesics in spacetime purely geometrically
and independent of the theory of gravity under consideration. Here,
$\theta$ is the expansion term describing how geodesics focus or
defocus; $\sigma_{ab}\sigma^{ab}$ is the shear which describes how,
e.g., a circular configuration of geodesics changes shape into, say,
an ellipse; $\omega_{ab}\omega^{ab}$ is the vorticity term; $R_{ab}$
is the Ricci tensor; and $U^{a}$ is the tangent vector to the geodesics.
Note that, due the sign of the expansion, shear, and the Ricci term,
they all contribute to focusing, while the vorticity terms leads to
defocusing. 

In our case, since we consider the model in vacuum, $R_{ab}=0$. Also,
in general in Kantowski-Scahs models, the vorticity term is only nonvanishing
if one considers metric perturbations \cite{Collins:1977fg}. Hence, $\omega_{ab}\omega^{ab}=0$
in our model, too. This reduces the Raychaudhuri equation for our analysis
to
\begin{equation}
\frac{d\theta}{d\tau}=-\frac{1}{3}\theta^{2}-\sigma_{ab}\sigma^{ab}.\label{eq:Ray-1}
\end{equation}
One can show that the above implies the convergence of
geodesics to conjugate points at a finite proper time $\tau_{0}<3/|\theta_{0}|,$
where $\theta_{0}$ is the starting expansion \cite{Raychaudhuri:1953yv}. 
This is due to the
fact that the right-hand side of the above is negative, which in turn is a direct
consequence of the universal and attractive nature of gravity. This
also reiterates the ``inevitability'' of geodesics focusing and
the consequent singularity theorems. 

\begin{figure}
\begin{centering}
\includegraphics[scale=0.7]{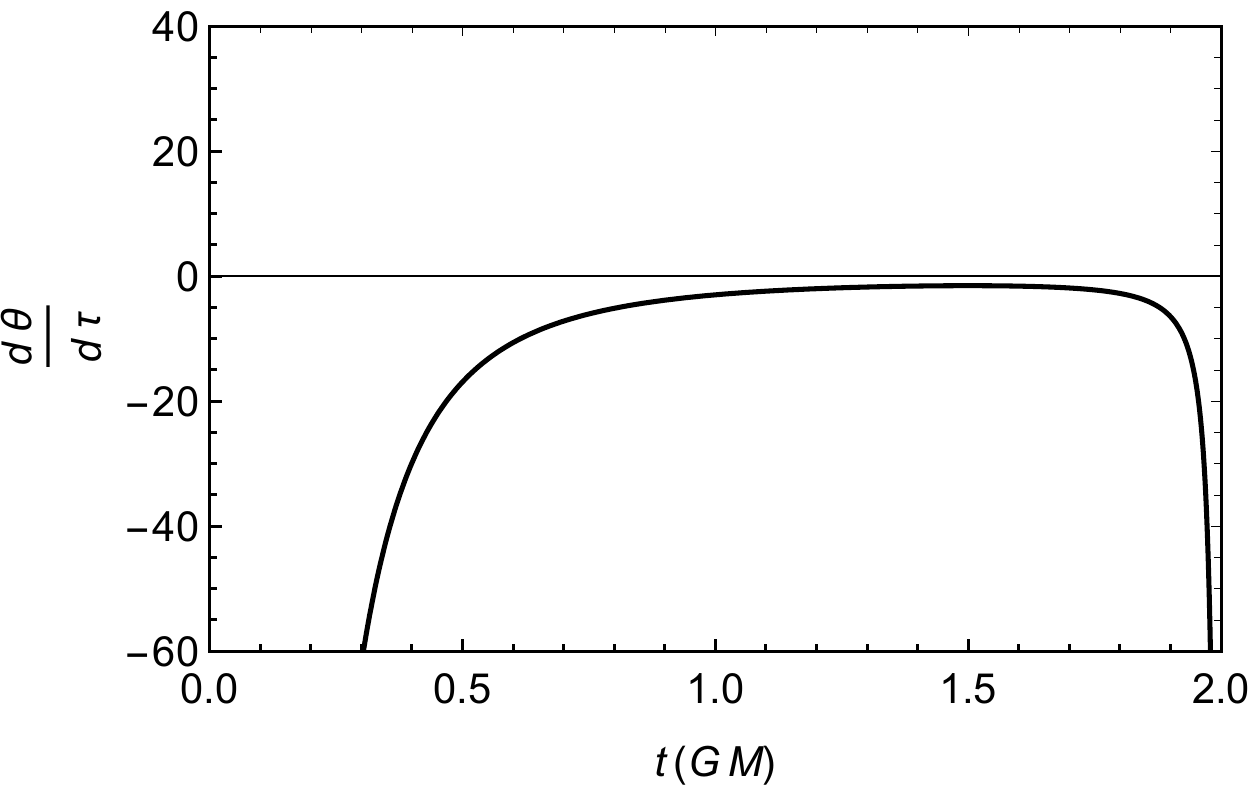}
\par\end{centering}
\caption{The right-hand side of the Raychaudhuri equation as a function of
Schwarzschild time $t$. At both the classical singularity $t\to0$
and at the horizon $t=2GM$, $\frac{d\theta}{d\tau}$ diverges. The
former is due to a physical singularity while the latter happens because
of the choice of coordinate system. To draw this plot, we have set
$M=1,\,G=1$. \label{fig:RE-cls-t}}
\end{figure}

In order to adapt the Raychaudhuri equation to the current LQG formalism,
we write the quantities $\theta$ and $\sigma^{2}=\sigma_{ab}\sigma^{ab}$
appearing on the right-hand side of (\ref{eq:Ray-1}) in terms of
the canonical variables as \cite{Corichi:2015xia}
\begin{align}
\theta= & \frac{\dot{p}_{b}}{Np_{b}}+\frac{\dot{p}_{c}}{2Np_{c}},\label{eq:expansion}\\
\sigma^{2}= & \frac{2}{3}\left(-\frac{\dot{p}_{b}}{Np_{b}}+\frac{\dot{p}_{c}}{Np_{c}}\right)^{2}.\label{eq:shear}
\end{align}
Replacing these in (\ref{eq:Ray-1}) and using the equations of motion
(\ref{eq:EoM-cls-b})--(\ref{eq:EoM-cls-pc}) one obtains
\begin{equation}
\frac{d\theta}{d\tau}=-\frac{1}{2p_{c}}\left(1+\frac{9b^{2}}{2\gamma^{2}}+\frac{\gamma^{2}}{2b^{2}}\right).\label{eq:Class-RE}
\end{equation}
As expected, the right hand side is negative (since $p_{c}>0$) and
diverges at the singularity given the behavior of canonical variables
in Fig. \ref{fig:class-var-behv}. This can also be seen from the
expression (\ref{eq:Class-RE}) written in terms of $t$, by using
the solutions (\ref{eq:sol-cls-b})-(\ref{eq:sol-cls-pc}) and the
lapse (\ref{eq:Sch-corresp-1}) in (\ref{eq:Class-RE}) to get
\begin{equation}
\frac{d\theta}{d\tau}=\frac{-2t^{2}+8GMt-9G^{2}M^{2}}{t^{\frac{5}{2}}\left(2GM-t\right)^{\frac{3}{2}}}.\label{eq:RE-RHS-t-class}
\end{equation}
The behavior of $\frac{d\theta}{d\tau}$ from (\ref{eq:RE-RHS-t-class})
is presented in Fig. (\ref{fig:RE-cls-t}). This figure confirms that
$\frac{d\theta}{d\tau}$ diverges at the classical singularity, pointing
to an infinite focusing of geodesics at that region. All of these
observations are well known. In what follows, we show that the quantum
effects modify this behavior particularly close to the classical singularity.

\section{Effective dynamics and Raychaudhuri equation\label{sec:Effective-Dynamics}}

The effective behavior of the interior of the Schwarzschild black
hole can be deduced from its effective Hamiltonian (constraint). There
are various ways to obtain such an effective Hamiltonian from the
classical one. Usually one first obtains the quantum Hamiltonian constraint.
This is done by first writing the full Hamiltonian (\ref{eq:Full-H-gr-class}),
particularly the curvature term(s), in terms of holonomies $h_{x},\,h_{\theta},\,h_{\phi}$
along edges in the radial and angular direction, and fluxes, instead
of the connection and the triad \cite{Ashtekar:2005qt,Corichi:2015xia,Chiou:2008nm,Chiou:2008eg}.
The main reason to do so is that in loop quantum gravity, the connection
is not a well-defined operator on the Hilbert space of the theory
\cite{Thiemann:2007pyv}. 

One then represents the holonomies and fluxes, and thus the Hamiltonian
constraint as an operator on a suitable Hilbert space \cite{Thiemann:2007pyv}.
Note that, due to the nonexistence of connection on the Hilbert space,
such a representation is unitarily inequivalent to the usual Schrodinger
representation \cite{Thiemann:2007pyv}. As a consequence, one obtains
distinct physical results compared to the Schrodinger representation.
Another property of LQG representation is that there exist no infinitesimal
diffeomorphisms, and one only has access to finite diffeomorphisms.
This leads to the discretization of space. For finite dimensional
systems, such a representation is isomorphic to the polymer representation.
This is a representation in which some of the operators are not weakly
continuous in their parameters \cite{Morales-Tecotl:2016ijb,Corichi:2007tf,Flores-Gonzalez:2013zuk,Tecotl:2015cya}.
In this case, the unitary inequivalence to the Schrodiner representation
follows directly from the Stone-von Neumann theorem \cite{StoneNeumann}.
Due to the existence of only finite transformations generated by some
operators, minimal scales appear in the theory, which then leads to
the discretization of some elements of the theory depending on what
operators exhibit only finite transformations. Usually, these minimal
scales are denoted by $\mu$ as we will see below.

After obtaining the quantum Hamiltonian as mentioned above, one finds an effective Hamiltonian by either using a path integral approach, or by
acting the quantum Hamiltonian on states peaked around some classical solutions \cite{Ashtekar:2002sn,Corichi:2007tf,Morales-Tecotl:2016ijb,Tecotl:2015cya,Flores-Gonzalez:2013zuk,Morales-Tecotl:2016dma,Morales-Tecotl:2018ugi}.
These methods will lead to an effective Hamiltonian that can also
be heuristically obtained by replacing 
\begin{align}
b\to & \frac{\sin\left(\mu_{b}b\right)}{\mu_{b}},\label{eq:b-to-sinb}\\
c\to & \frac{\sin\left(\mu_{c}c\right)}{\mu_{c}}\label{eq:c-to-sinc}
\end{align}
in the classical Hamiltonian.

The free parameters $\mu_{b},\,\mu_{c}$ are the minimum scales associated
with the radial and angular directions \cite{Ashtekar:2005qt,Corichi:2015xia,Chiou:2008nm,Chiou:2008eg}.
In LQG, there exist two general schemes regarding these $\mu$ parameters.
In one, called the $\mu_{0}$ scheme, $\mu$ parameters are considered
to be constant \cite{Ashtekar:2005qt,Modesto:2005zm,Modesto:2008im,Campiglia:2007pr}.
Applying such a scheme to isotropic and Bianchi-I cosmological models,
however, has shown to lead to incorrect semiclassical limit. To remedy
this and other issues regarding the appearance of large quantum effects
at the horizon or dependence of physical quantities on fiducial variables,
new schemes referred to as the $\bar{\mu}$ scheme or ``improved dynamics''
have been proposed in which $\mu$ parameters depend on canonical
variables \cite{Bohmer:2007wi,Chiou:2008nm,Chiou:2008eg,Joe:2014tca}.
This scheme is itself divided into various different ways of expressing
the dependence of $\mu$ parameters on canonical variables. In addition,
new $\mu_{0}$ schemes have also been put forward (e.g., Refs. \cite{Corichi:2015xia,Olmedo:2017lvt})
with the intent of resolving the aforementioned issues. 

In case of the Schwarzschild interior due to lack of matter content,
it is not clear which scheme does not lead to the correct semiclassical
limit. Hence for completeness, in this paper, we will study the modifications
to the Raychaudhuri equation in the constant $\mu$ scheme, which
here we call the $\mathring{\mu}$ scheme, as well as in two of the
most common improved schemes, which we denote by $\bar{\mu}$ and $\bar{\mu}^{\prime}$
schemes.

Applying any of the methods of deriving an effective Hamiltonian or
simply replacing (\ref{eq:b-to-sinb}) and (\ref{eq:c-to-sinc}) into
the classical Hamiltonian (\ref{eq:H-const-class}), one obtains an
effective Hamiltonian constraint,
\begin{equation}
H_{\textrm{eff}}^{(N)}=-\frac{N}{2G\gamma^{2}}\left[\left(\frac{\sin^{2}\left(\mu_{b}b\right)}{\mu_{b}^{2}}+\gamma^{2}\right)\frac{p_{b}}{\sqrt{p_{c}}}+2\frac{\sin\left(\mu_{b}b\right)}{\mu_{b}}\frac{\sin\left(\mu_{c}c\right)}{\mu_{c}}\sqrt{p_{c}}\right].\label{eq:H-eff-gen}
\end{equation}
In order to be able to find the deviations from the classical behavior,
we need to use the same lapse as we did in the classical part. Under
(\ref{eq:b-to-sinb}), this lapse (\ref{eq:lapsNT}) becomes
\begin{equation}
N=\frac{\gamma\mu_{b}\sqrt{p_{c}}}{\sin\left(\mu_{b}b\right)}.\label{eq:laps-eff}
\end{equation}
Using this in (\ref{eq:H-eff-gen}) yields
\begin{equation}
H_{\textrm{eff}}=-\frac{1}{2\gamma G}\left[p_{b}\left[\frac{\sin\left(\mu_{b}b\right)}{\mu_{b}}+\gamma^{2}\frac{\mu_{b}}{\sin\left(\mu_{b}b\right)}\right]+2p_{c}\frac{\sin\left(\mu_{c}c\right)}{\mu_{c}}\right].\label{eq:H-eff-spec}
\end{equation}
Note that both (\ref{eq:H-eff-gen}) and (\ref{eq:H-eff-spec}) reduce
to their classical counterparts (\ref{eq:H-const-class}) and (\ref{eq:H-const-cls-gauged})
respectively, as is expected. 

\subsection{$\mathring{\mu}$ scheme\label{subsec:mu0-scheme}}

\begin{figure}
\begin{centering}
\includegraphics[scale=0.7]{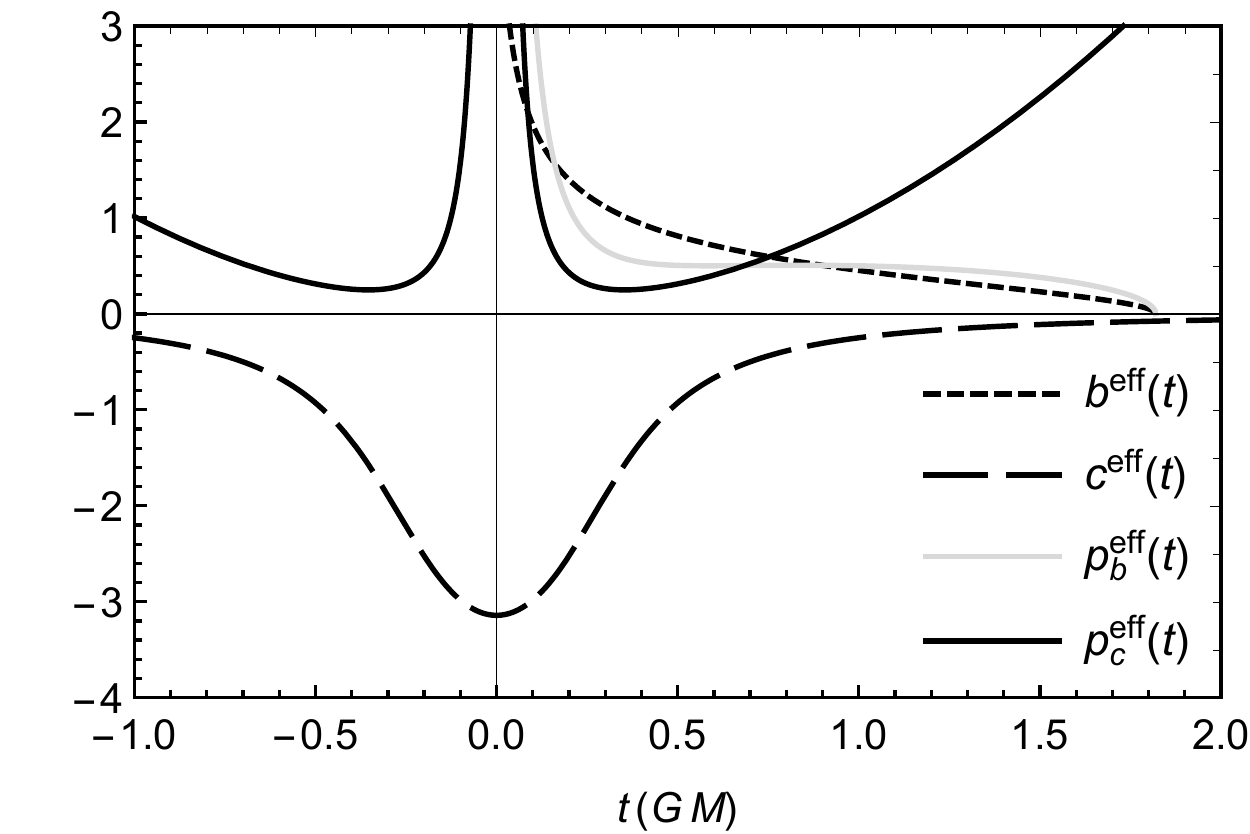}
\par\end{centering}
\caption{The behavior of canonical variables as a function of Schwarzschild
time $t$. We have chosen the positive sign for $b$ and negative
sign for $c$. The figure is plotted using $\gamma=0.5,\,M=1,\,G=1,\,L_{0}=1$
and $\mathring{\mu}_{b}=0.5=\mathring{\mu}_{c}$. Notice the bounce
in $p_{b},\,p_{c}$ and also in $c$. \label{fig:EoM-eff-mu0}}
\end{figure}

As mentioned before, in this scheme, one assumes that the polymer or
minimal scales $\mathring{\mu}_{b},\,\mathring{\mu}_{c}$ are constants.
Hence, the equations of motion corresponding to (\ref{eq:H-eff-spec})
become
\begin{align}
\frac{db}{dT}= & \left\{ b,H_{\textrm{eff}}\right\} =-\frac{1}{2}\left[\frac{\sin\left(\mathring{\mu}_{b}b\right)}{\mathring{\mu}_{b}}+\gamma^{2}\frac{\mathring{\mu}_{b}}{\sin\left(\mathring{\mu}_{b}b\right)}\right],\\
\frac{dp_{b}}{dT}= & \left\{ p_{b},H_{\textrm{eff}}\right\} =\frac{1}{2}p_{b}\cos\left(\mathring{\mu}_{b}b\right)\left[1-\gamma^{2}\frac{\mathring{\mu}_{b}^{2}}{\sin^{2}\left(\mathring{\mu}_{b}b\right)}\right],\label{eq:pb-diff-mu0}\\
\frac{dc}{dT}= & \left\{ c,H_{\textrm{eff}}\right\} =-2\frac{\sin\left(\mathring{\mu}_{c}c\right)}{\mathring{\mu}_{c}},\\
\frac{dp_{c}}{dT}= & \left\{ p_{c},H_{\textrm{eff}}\right\} =2p_{c}\cos\left(\mathring{\mu}_{c}c\right).\label{eq:pc-diff-mu0}
\end{align}
Notice that the $\mathring{\mu}_{b}\to0$ and $\mathring{\mu}_{c}\to0$
limit of these equations corresponds to the classical equations of
motion (\ref{eq:EoM-cls-b})--(\ref{eq:EoM-cls-pc}). The solutions
to these equations in terms of the Schwarzschild time $t$ (after
a transformation $T=\ln(t)$) and choosing suitable initial conditions,
are given by
\begin{align}
b(t)= & \frac{\cos^{-1}\left[\sqrt{1+\gamma^{2}\mathring{\mu}_{b}^{2}}\tanh\left(\sqrt{1+\gamma^{2}\mathring{\mu}_{b}^{2}}\ln\left[\frac{2\sqrt{\frac{t}{2M}}}{\gamma\mathring{\mu}_{b}}\right]\right)\right]}{\mathring{\mu}_{b}},\label{eq:bt-eff-mu0}\\
p_{b}(t)= & \frac{\gamma\mathring{\mu}_{b}L_{0}M\left(\frac{\gamma^{2}\mathring{\mu}_{c}^{2}L_{0}^{2}M^{2}}{4t^{2}}+t^{2}\right)\sqrt{1-\left(1+\gamma^{2}\mathring{\mu}_{b}^{2}\right)\tanh^{2}\left(\sqrt{\gamma^{2}\mathring{\mu}_{b}^{2}+1}\ln\left[\frac{2\sqrt{\frac{t}{2M}}}{\gamma\mathring{\mu}_{b}}\right]\right)}}{t^{2}\sqrt{\frac{\gamma^{2}\mathring{\mu}_{c}^{2}L_{0}^{2}M^{2}}{4t^{4}}+1}\left(\gamma^{2}\mathring{\mu}_{b}^{2}-\left(1+\gamma^{2}\mathring{\mu}_{b}^{2}\right)\tanh^{2}\left(\sqrt{1+\gamma^{2}\mathring{\mu}_{b}^{2}}\ln\left[\frac{2\sqrt{\frac{t}{2M}}}{\gamma\mathring{\mu}_{b}}\right]\right)+1\right)},\label{eq:eq:pbt-eff-mu0}\\
c(t)= & -\frac{\tan^{-1}\left(\frac{\gamma\mathring{\mu}_{c}L_{0}M}{2t^{2}}\right)}{\mathring{\mu}_{c}},\label{eq:ct-eff-mu0}\\
p_{c}(t)= & \frac{\gamma^{2}\mathring{\mu}_{c}^{2}L_{0}^{2}M^{2}}{4t^{2}}+t^{2}.\label{eq:pct-eff-mu0}
\end{align}
The behavior of these solutions is plotted in Fig. \ref{fig:EoM-eff-mu0}.
It is seen that both $p_{b}$ and $p_{c}$ exhibit a bounce in this
effective regime. Particularly, there is a minimum radius-at-the-bounce
due to the existence of a minimum value for $p_{c}$. This leads to the
resolution of the classical singularity which can, e.g., be seen from
the fact that the Riemann invariants, which are proportional to $\frac{1}{p_{c}^{n}}$
with $n>0$, do not diverge anywhere inside the black hole. 

Replacing the effective solutions (\ref{eq:pb-diff-mu0}) and (\ref{eq:pc-diff-mu0})
into (\ref{eq:expansion}) and (\ref{eq:shear}) and using them in
the expression of the Raychaudhuri equation (\ref{eq:Ray-1}), one
obtains
\begin{align}
\frac{d\theta}{d\tau}= & \frac{1}{\gamma^{2}p_{c}}\frac{\sin^{2}\left(\mathring{\mu}_{b}b\right)}{\mathring{\mu}_{b}^{2}}\left[\cos\left(\mathring{\mu}_{b}b\right)\cos\left(\mathring{\mu}_{c}c\right)-\frac{\cos^{2}\left(\mathring{\mu}_{b}b\right)}{4}-3\cos^{2}\left(\mathring{\mu}_{c}c\right)\right]\nonumber \\
 & +\frac{\cos\left(\mathring{\mu}_{b}b\right)}{p_{c}}\left[\frac{\cos\left(\mathring{\mu}_{b}b\right)}{2}-\cos\left(\mathring{\mu}_{c}c\right)-\frac{\gamma^{2}}{4}\cos\left(\mathring{\mu}_{b}b\right)\frac{\mathring{\mu}_{b}^{2}}{\sin^{2}\left(\mathring{\mu}_{b}b\right)}\right].\label{eq:dtheta-dtau-mu0}
\end{align}
Before considering the full nonperturbative expression above, let
us look at its expansion up to the second order in $\mathring{\mu}_{b}$
and $\mathring{\mu}_{c}$,
\begin{equation}
\frac{d\theta}{d\tau}\approx-\frac{1}{2p_{c}}\left(1+\frac{9b^{2}}{2\gamma^{2}}+\frac{\gamma^{2}}{2b^{2}}\right)+\mathring{\mu}_{b}^{2}\frac{1}{2p_{c}}\left(\frac{b^{4}}{\gamma^{2}}+\frac{\gamma^{2}}{3}\right)+\mathring{\mu}_{c}^{2}\frac{c^{2}}{2p_{c}}\left(1+\frac{5b^{2}}{\gamma^{2}}\right).\label{eq:dtheta-dtau-mu0-pert}
\end{equation}
One can see that the first term on the right-hand side is the classical
expression (\ref{eq:Class-RE}) which is always negative and leads
to the divergence of classical expansion rate at the singularity,
i.e., infinite focusing. However, Eq. (\ref{eq:dtheta-dtau-mu0-pert})
now involves two additional effective terms proportional to $\mathring{\mu}_{b}^{2}$ and $\mathring{\mu}_{c}^{2}$, both of which are positive. Hence, the quantum corrections
up to the second order in polymer parameters contribute to defocusing,
which becomes particularly large close to the singularity. 

\begin{figure}
\begin{centering}
\includegraphics[scale=0.7]{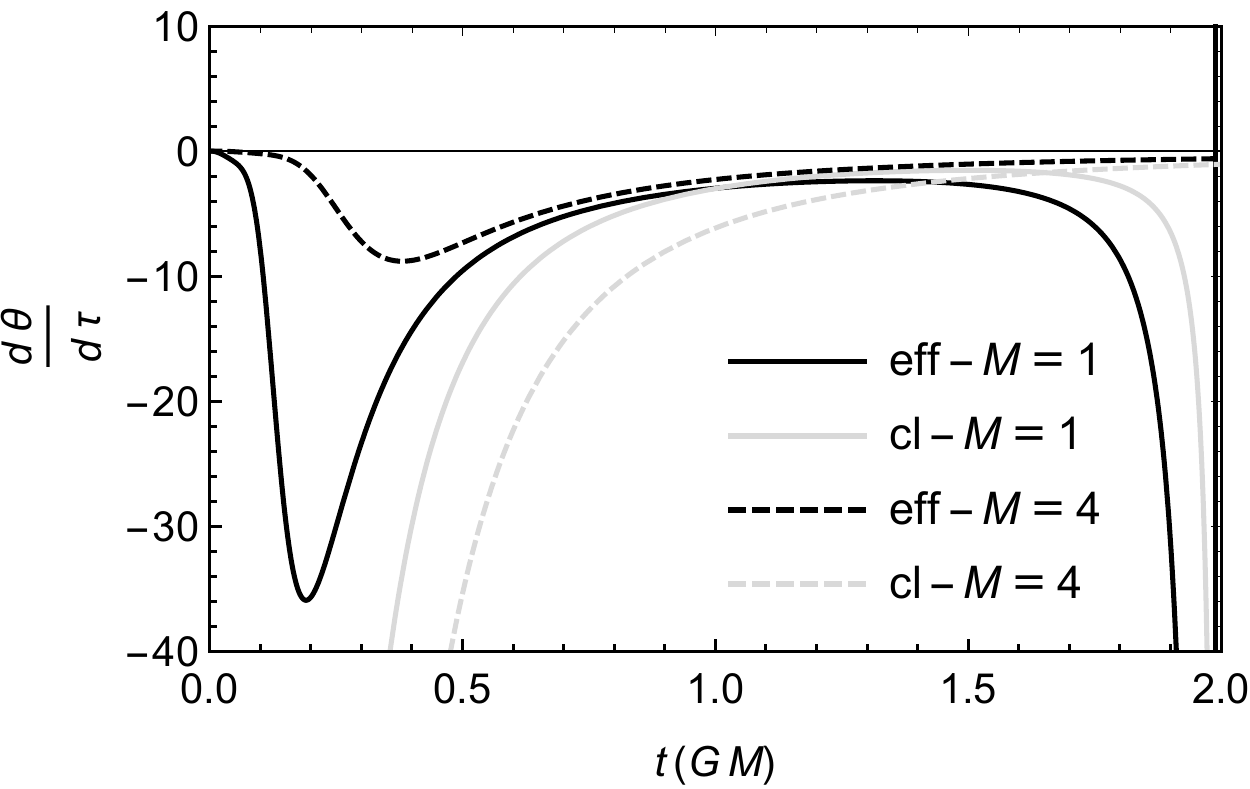}
\par\end{centering}
\caption{Plot of $\frac{d\theta}{d\tau}$ as a function of the Schwarzschild
time $t$, for two different masses in classical vs effective
regimes. The figure is plotted using $\gamma=0.5,\,G=1,\,L_{0}=1$,
and $\mathring{\mu}_{b}=0.08=\mathring{\mu}_{c}$. \label{fig:RE-eff-mu0}}
\end{figure}

This is in fact
confirmed by considering the full nonperturbative expression (\ref{eq:dtheta-dtau-mu0})
written in terms of the Schwarzschild time $t$ using the solutions
(\ref{eq:bt-eff-mu0})--(\ref{eq:pct-eff-mu0}). We do not present
this expression here since it is very lengthy. However, we have plotted
this expression in Fig. \ref{fig:RE-eff-mu0}. This plot includes
the classical versus the effective behavior of $\frac{d\theta}{d\tau}$
for two different masses. First, we can see that, while $\frac{d\theta}{d\tau}$
diverges at the classical singularity, signaling an infinite focusing
of geodesics there for both masses, the quantum effects actually reverse
this situation for the effective case. Consequently, $\frac{d\theta}{d\tau}$
bounces back from a finite negative value and goes to zero when we
approach the classical singularity region. This leads to the resolution
of the singularity in the effective theory. Second, note that this
happens earlier for a larger black hole.

\begin{figure}
\begin{centering}
\includegraphics[scale=0.35]{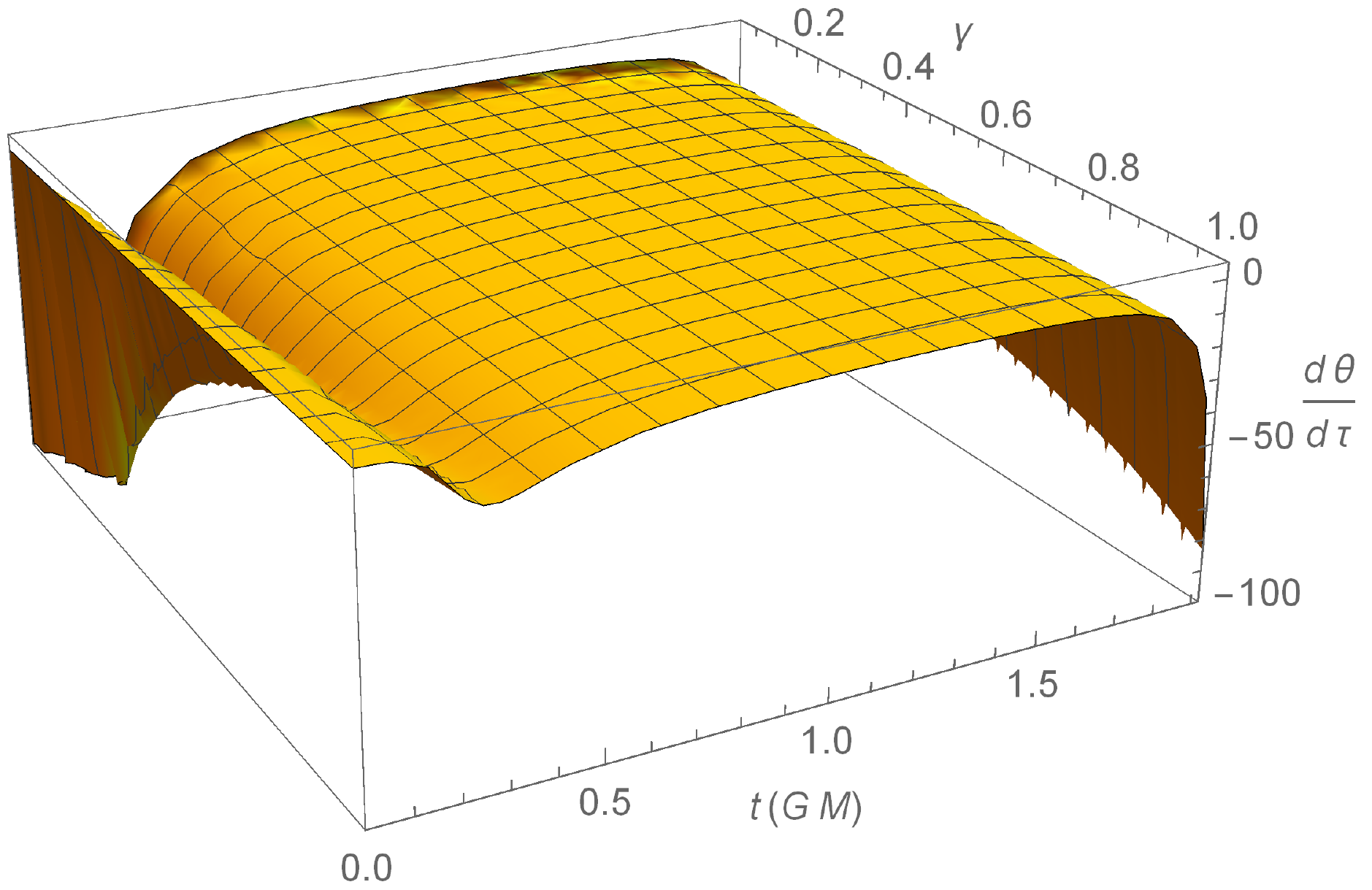}~~~~~~~~~~~~~~~~~\includegraphics[scale=0.35]{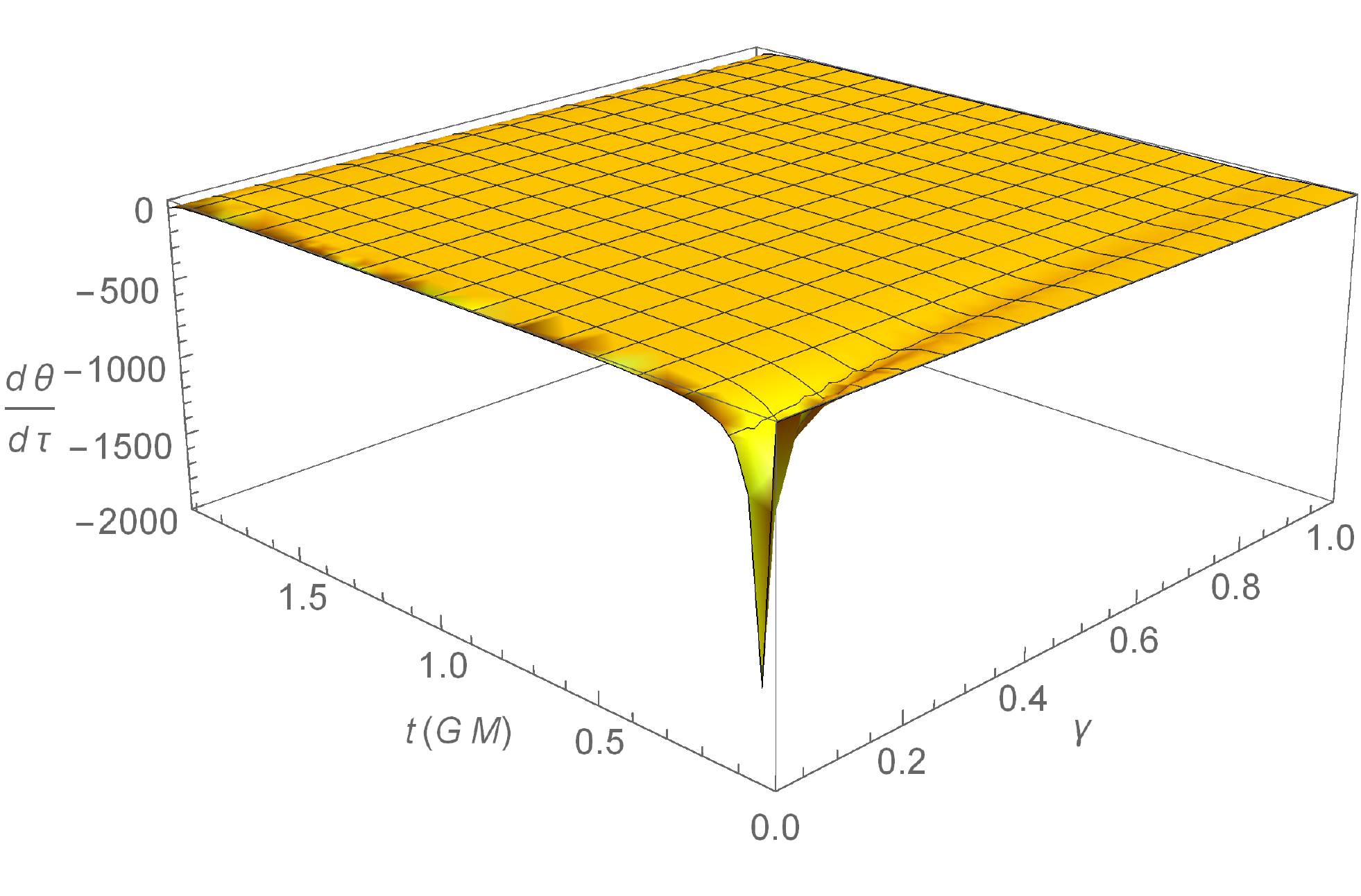}
\par\end{centering}
\caption{Left: plot of $\frac{d\theta}{d\tau}$ as a function of both $t$ and $\gamma$. Right: The same plot with the full interval of $\frac{d\theta}{d\tau}$ in picture. The values used here are $\,G=1,\,L_{0}=1$,
and $\mathring{\mu}_{b}=0.08=\mathring{\mu}_{c}$. \label{fig:RE-vs-pc-vs-gamma-Eff-mu0}}
\end{figure}

Looking back at the Raychaudhuri equation (\ref{eq:Ray-1}), we see
that both terms in that equation carry a negative sign, so, classically,
they both contribute to focusing of geodesics, and this focusing becomes
larger and larger with no other term being present to counter it.
However, the nonperturbative effective correction terms resulting
from loop quantization contribute to terms that have a positive sign
and only become significant when one gets close to the singularity
where quantum gravity effects should be significant. These effective
terms then take over and revert the focusing property of the classical
terms, so much so that they return the value of $\frac{d\theta}{d\tau}$
to zero at the region that used to be the classical singularity.

\begin{figure}
\begin{centering}
\includegraphics[scale=0.55]{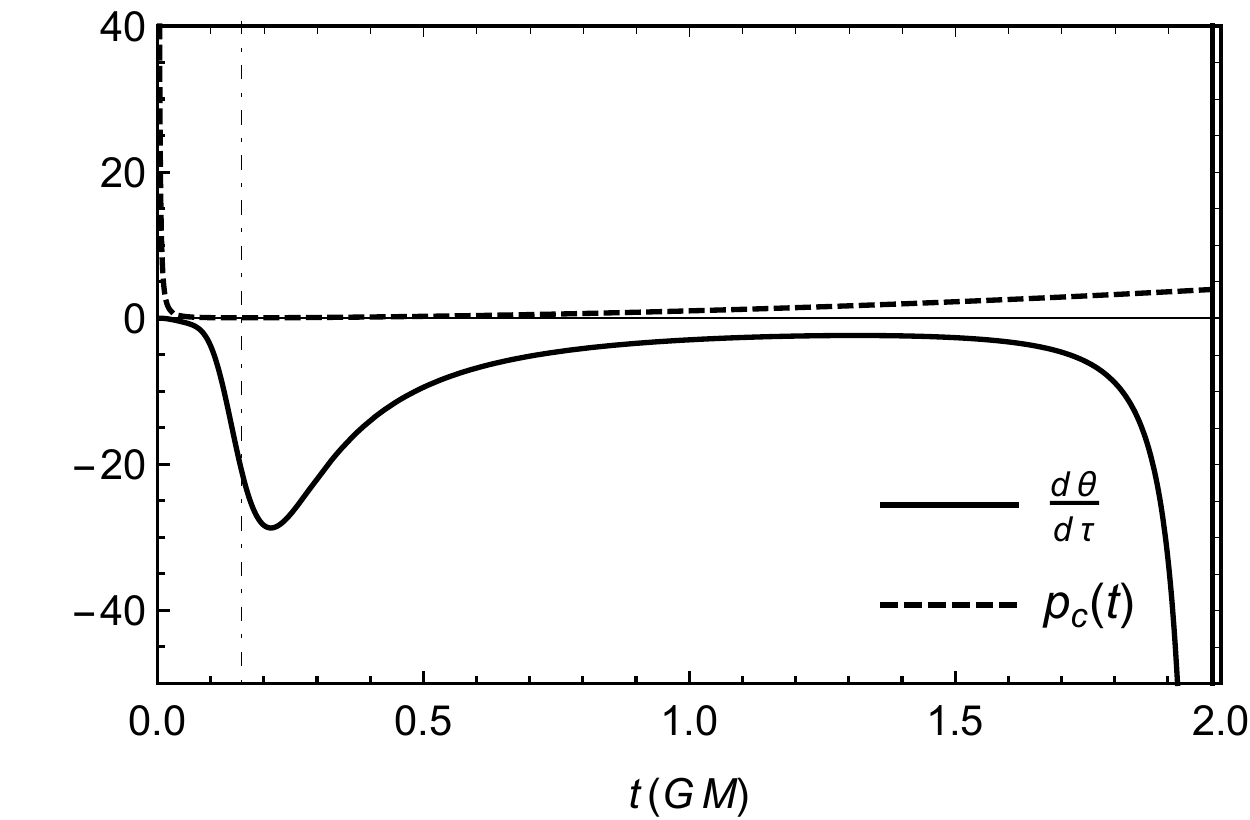}~~~~~~~~~~~~~~~~~\includegraphics[scale=0.55]{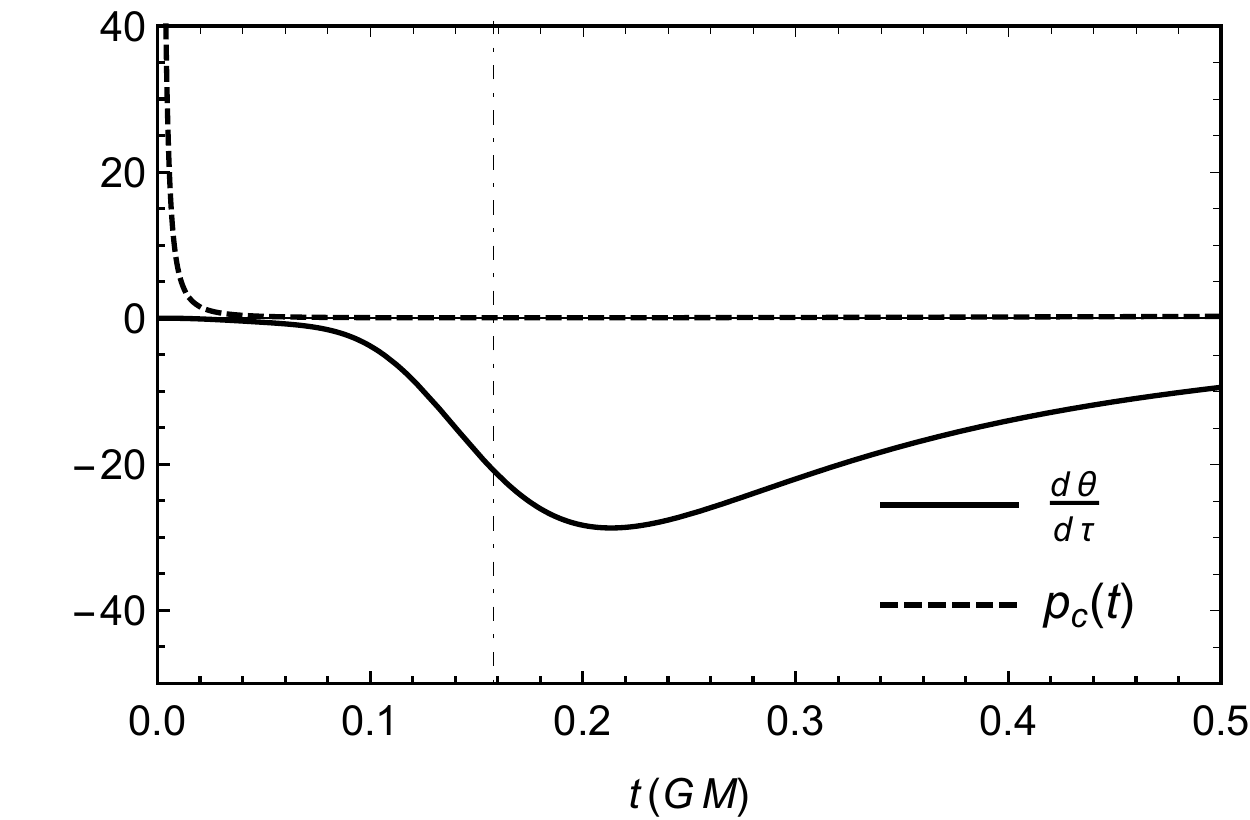}
\par\end{centering}
\caption{Left: plot of $\frac{d\theta}{d\tau}$ vs $p_{c}(t)$ as a function
of the Schwarzschild time $t$. Right: the close-up of $0\protect\leq t\protect\leq0.5GM$
portion of the left figure. The vertical dot-dashed line shows the
time $t\approx0.158GM$ when the minimum of $p_{c}^{\textrm{min}}=0.05$
happens in this case. These are plotted using $\gamma=0.5,\,G=1,\,L_{0}=1$,
and $\mathring{\mu}_{b}=0.1=\mathring{\mu}_{c}$. \label{fig:RE-vs-pc-eff-mu0}}
\end{figure}

To get more insight, we also have plotted $\frac{d\theta}{d\tau}$
against the (square of the) radius of 2-spheres, $p_{c}(t)$, in
Fig. \ref{fig:RE-vs-pc-eff-mu0}. From this figure, one can see that
both $p_{c}$ and $\frac{d\theta}{d\tau}$ bounce back. However, $\frac{d\theta}{d\tau}$
starts bouncing back much earlier than $p_{c}$ and ``knows'' about
the ``repulsive'' quantum gravity effects much more earlier than
$p_{c}$ does. By the time $p_{c}$ starts bouncing back, $\frac{d\theta}{d\tau}$
is already bouncing toward zero value.

Finally, in Fig. \ref{fig:RE-vs-pc-vs-gamma-Eff-mu0}, we have plotted $\frac{d\theta}{d\tau}$ as a function of both $t$ and the Barbero-Immirzi parameter $\gamma$. It is seen that with given $\mathring{\mu}_b$ and $\mathring{\mu}_c$, decrease in value of $\gamma$ deepens the minimum of $\frac{d\theta}{d\tau}$ but in $\mathring{\mu}$ scheme there always will be a bounce anyway while $\gamma>0$.

\subsection{$\bar{\mu}$ scheme\label{subsec:mubar-1-scheme}}

\begin{figure}
\begin{centering}
\includegraphics[scale=0.55]{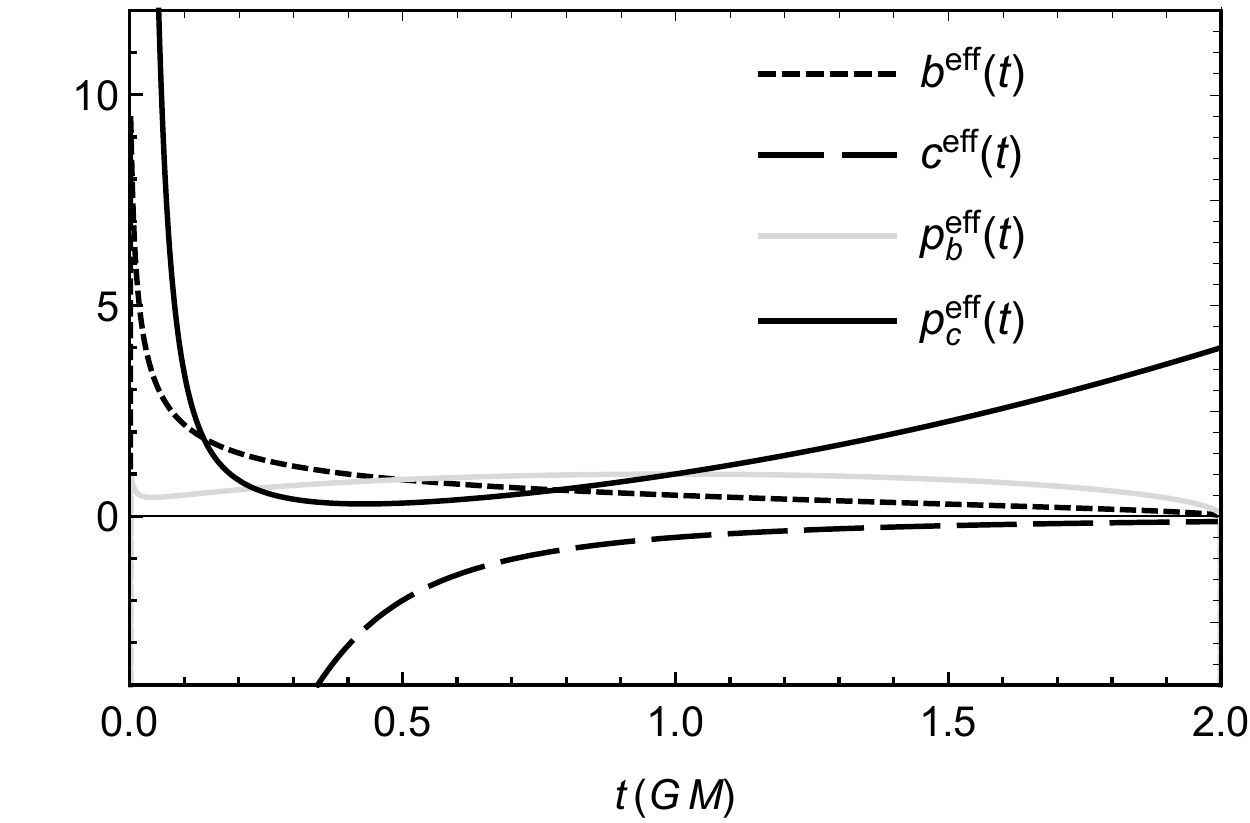}
\par\end{centering}
\caption{The behavior of canonical variables as a function of Schwarzschild
time $t$ in the $\bar{\mu}$ scheme. We have chosen the positive sign
for $b$ and negative sign for $c$. The figure is plotted using $\gamma=0.5,\,M=1,\,G=1,\,L_{0}=1$,
and $\Delta=0.1$. \label{fig:All-sol-bar-t}}
\end{figure}

In this scheme $\bar{\mu}_{b},\,\bar{\mu}_{c}$ are assumed to depend
on the triad components as
\begin{align}
\bar{\mu}_{b}= & \sqrt{\frac{\Delta}{p_{b}}},\label{eq:mu-bar-b}\\
\bar{\mu}_{c}= & \sqrt{\frac{\Delta}{p_{c}}}.\label{eq:mu-bar-c}
\end{align}
Using the same lapse as (\ref{eq:laps-eff}) but keeping in mind the
above dependence of $\bar{\mu}_{b},\,\bar{\mu}_{c}$, one can easily
obtain the equations of motion as
\begin{align}
\frac{db}{dT}= & \frac{1}{4}\left(b\cos\left(\bar{\mu}_{b}b\right)-3\frac{\sin\left(\bar{\mu}_{b}b\right)}{\bar{\mu}_{b}}-\gamma^{2}\frac{\bar{\mu}_{b}}{\sin\left(\bar{\mu}_{b}b\right)}\left[1+b\cos\left(\bar{\mu}_{b}b\right)\frac{\bar{\mu}_{b}}{\sin\left(\bar{\mu}_{b}b\right)}\right]\right),\\
\frac{dp_{b}}{dT}= & \frac{1}{2}p_{b}\cos\left(\bar{\mu}_{b}b\right)\left[1-\gamma^{2}\frac{\bar{\mu}_{b}^{2}}{\sin^{2}\left(\bar{\mu}_{b}b\right)}\right],\\
\frac{dc}{dT}= & c\cos\left(\bar{\mu}_{c}c\right)-3\frac{\sin\left(\bar{\mu}_{c}c\right)}{\bar{\mu}_{c}},\\
\frac{dp_{c}}{dT}= & 2p_{c}\cos\left(\bar{\mu}_{c}c\right).
\end{align}
These solutions are plotted in Fig. \ref{fig:All-sol-bar-t}. To obtain
them we have solved the system numerically and used the same initial
conditions very close to the horizon as the classical solutions. From
Fig. \ref{fig:All-sol-bar-t} one can see that $p_{c}$ still retains
a certain minimum value and bounces after reaching this value. Hence,
again, none of the Riemann invariants will diverge. 

The full Raychaudhuri equation now becomes
\begin{align}
\frac{d\theta}{d\tau}= & \frac{1}{\gamma^{2}p_{c}}\frac{\sin^{2}\left(\bar{\mu}_{b}b\right)}{\bar{\mu}_{b}^{2}}\left[\cos\left(\bar{\mu}_{b}b\right)\cos\left(\bar{\mu}_{c}c\right)-\frac{\cos^{2}\left(\bar{\mu}_{b}b\right)}{4}-3\cos^{2}\left(\bar{\mu}_{c}c\right)\right]\nonumber \\
 & +\frac{\cos\left(\bar{\mu}_{b}b\right)}{p_{c}}\left[\frac{\cos\left(\bar{\mu}_{b}b\right)}{2}-\cos\left(\bar{\mu}_{c}c\right)-\frac{\gamma^{2}}{4}\cos\left(\bar{\mu}_{b}b\right)\frac{\bar{\mu}_{b}^{2}}{\sin^{2}\left(\bar{\mu}_{b}b\right)}\right].\label{eq:dtheta-dtau-mu-bar-1}
\end{align}
This looks similar in form to (\ref{eq:dtheta-dtau-mu0}), but we
should keep in mind that the polymer parameters here depend on canonical
variables as in (\ref{eq:mu-bar-b}) and (\ref{eq:mu-bar-c}). As
in the previous case, let us consider the perturbative expansion of
this expression before considering the full nonperturbative version.
Up to first order in $\Delta$ (which can be considered as the second
order in $\bar{\mu}$ scales), we get
\begin{equation}
\frac{d\theta}{d\tau}\approx-\frac{1}{2p_{c}}\left(1+\frac{9b^{2}}{2\gamma^{2}}+\frac{\gamma^{2}}{2b^{2}}\right)+\frac{\Delta}{p_{c}}\left[\frac{1}{6p_{b}}\left(\frac{3b^{4}}{\gamma^{2}}+\gamma^{2}\right)+\frac{c^{2}}{2p_{c}}\left(1+\frac{5b^{2}}{\gamma^{2}}\right)\right].
\end{equation}
Once again, the first term on the right-hand side is the classical
expression of the Raychaudhuri equation (\ref{eq:Class-RE}), which
contributes to infinite focusing at the singularity. However, all
the correction terms are positive and once again these terms contribute to defocusing which
becomes significant close to the singularity.

\begin{figure}
\begin{centering}
\includegraphics[scale=0.55]{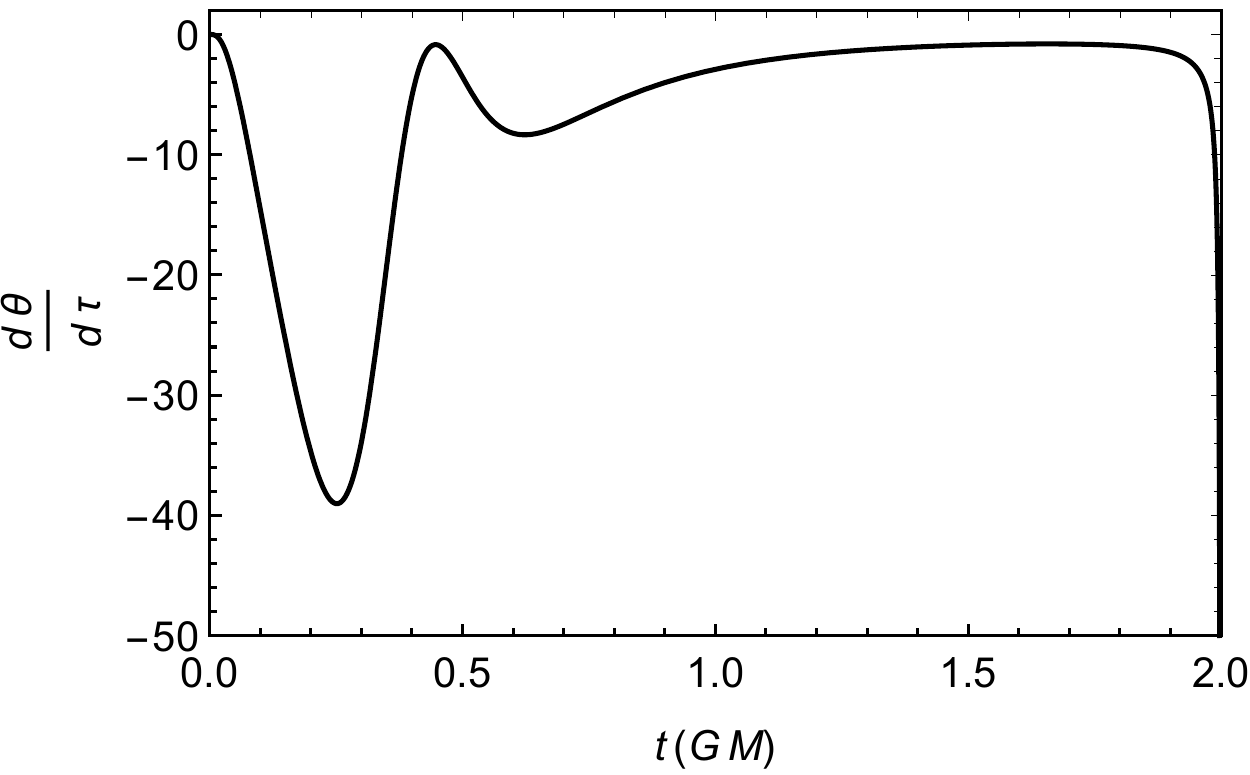}~~~~~~~~~~~~~~~~~\includegraphics[scale=0.55]{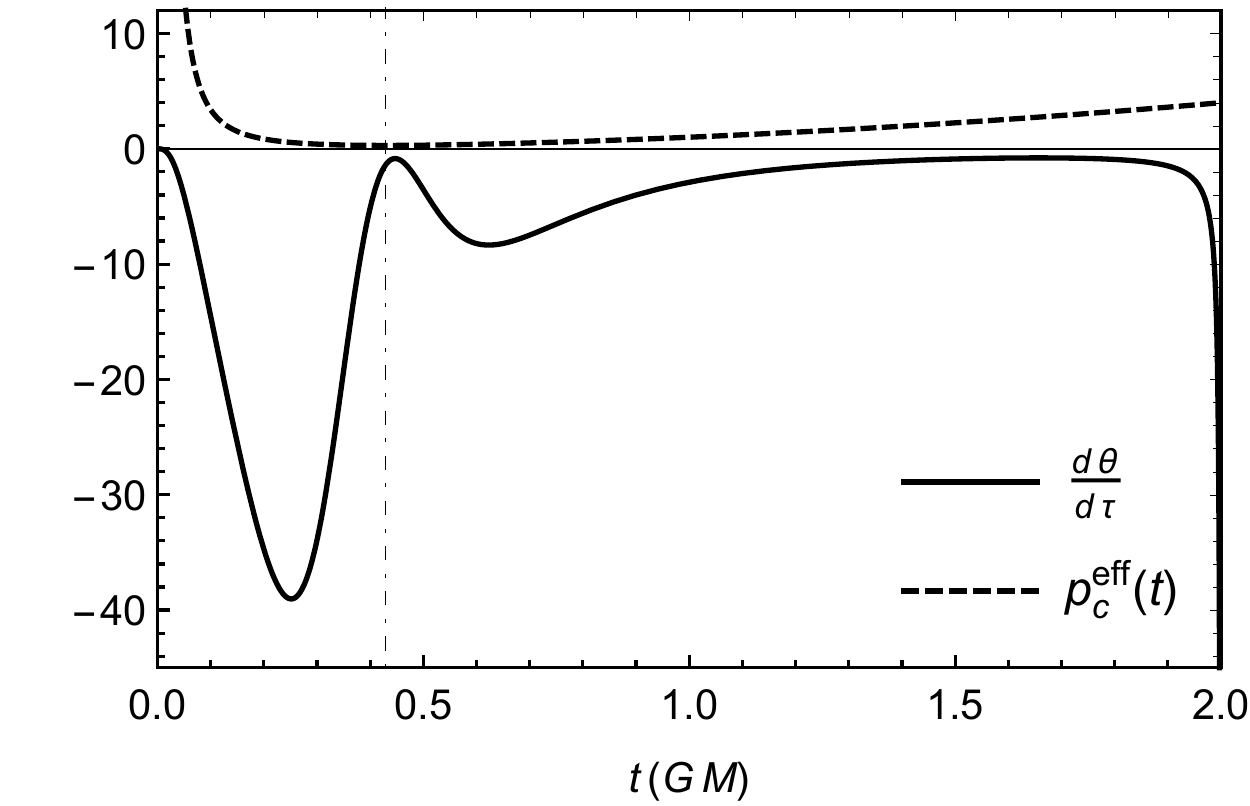}
\par\end{centering}
\caption{Left: Raychaudhuri equation in the $\bar{\mu}$ scheme. Right: Raychaudhuri
equation vs $p_{c}$. The vertical dot-dashed line at $t\approx0.43GM$
is the position of the bounce of $p_{c}$ where its minimum $p_{c}^{\textrm{min}}=0.29$
happens in this case. The figure is plotted using $\gamma=0.5,\,M=1,\,G=1,\,L_{0}=1$,
and $\Delta=0.1$. \label{fig:RE-bar}}
\end{figure}

The full nonperturbative form the modified Raychaudhuri equation in
terms of $t$ is plotted in Fig. \ref{fig:RE-bar}. We see that, approaching
from the horizon to where  the classical singularity used to be, an
initial bump or bounce in encountered, followed by a more pronounced
bounce closer to where the singularity used to be. Once again, the
quantum corrections become dominant close to the singularity and turn
back the $\frac{d\theta}{d\tau}$ such that at $t\to0$ no focusing
happens at all. Furthermore, from the right plot in Fig. \ref{fig:RE-bar},
we see that the first bounce in the Raychaudhuri equation happens
much earlier than the bounce in $p_{c}$.

\begin{figure}
\begin{centering}
\includegraphics[scale=0.5]{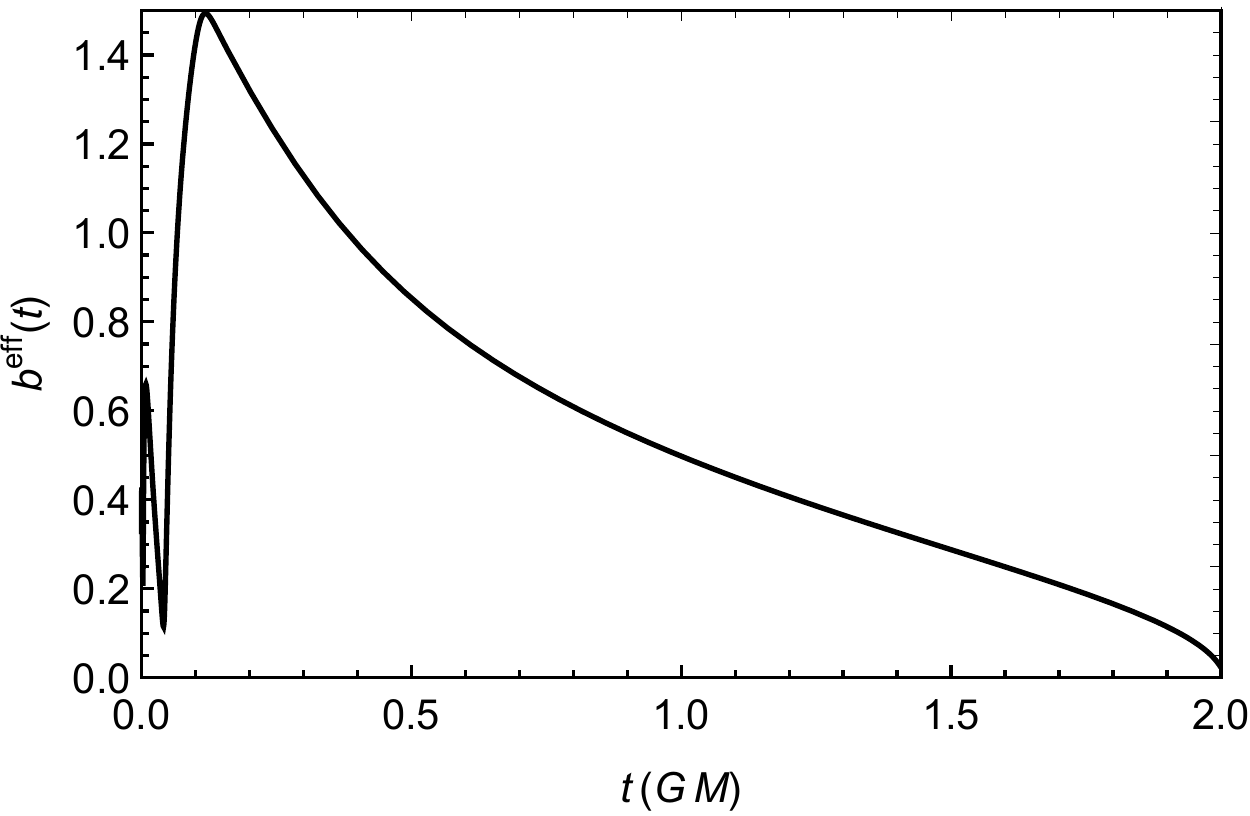}~~~~~~~~\includegraphics[scale=0.5]{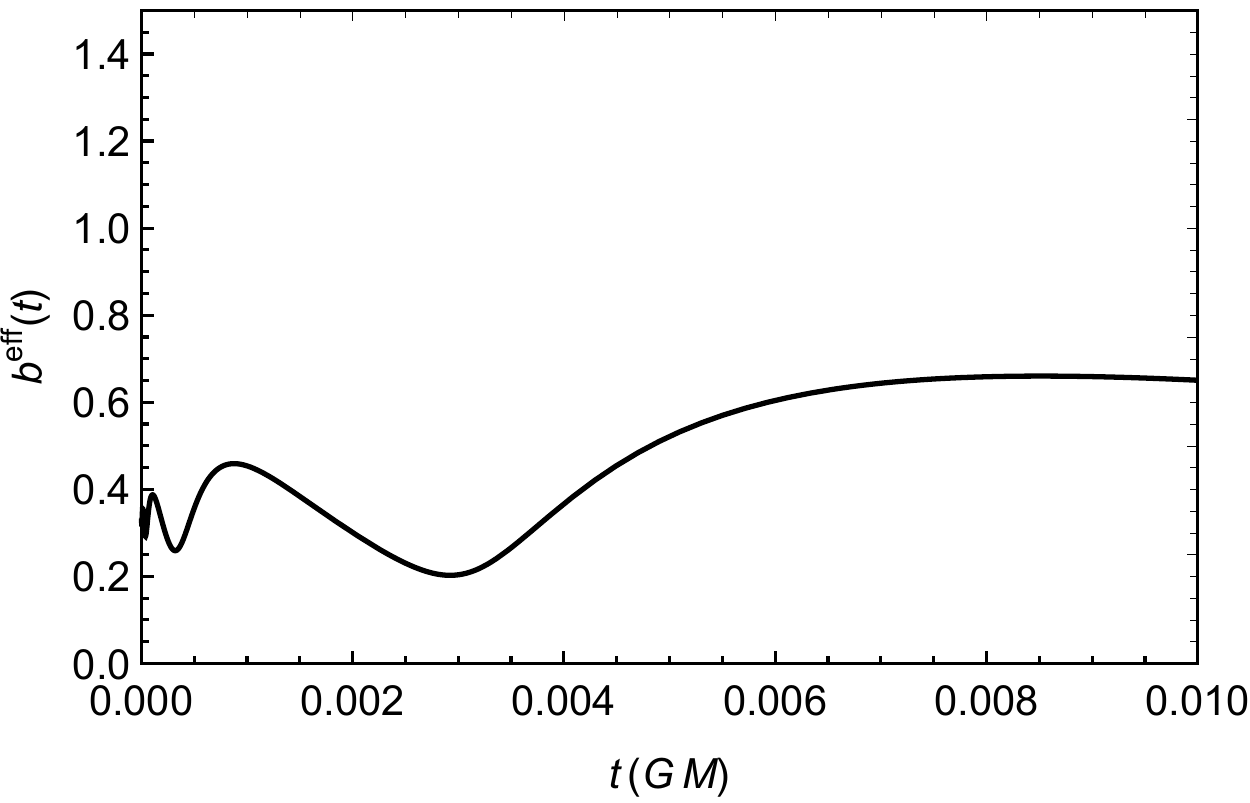}
\par\end{centering}
\begin{centering}
\vspace{10pt}
\includegraphics[scale=0.5]{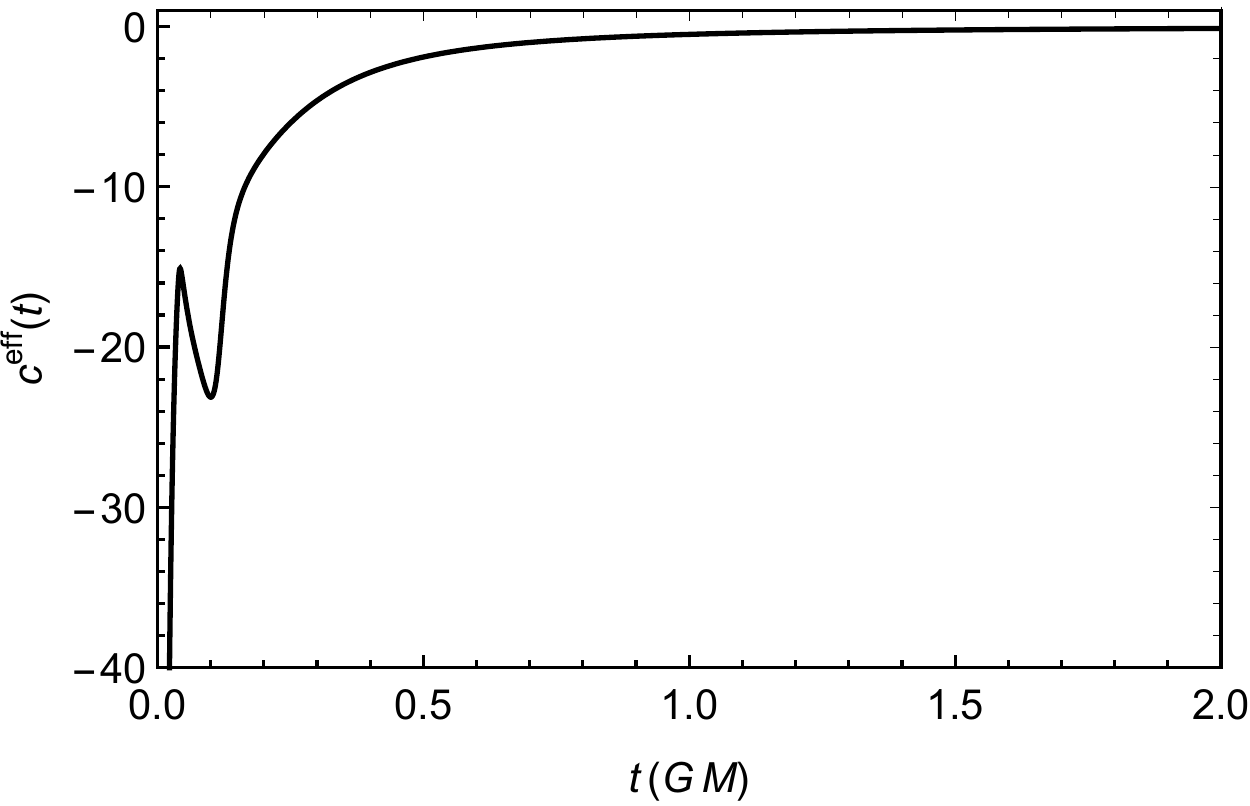}~~~~~~~~\includegraphics[scale=0.5]{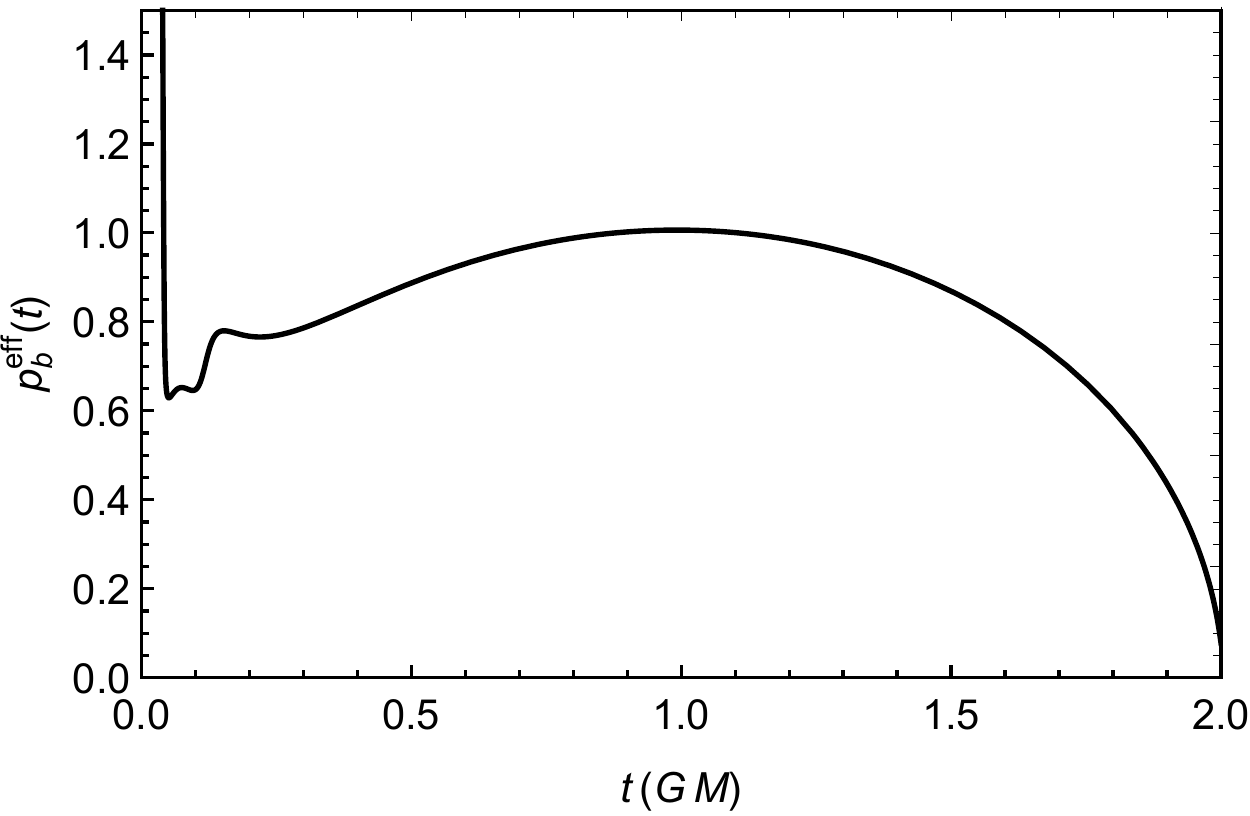}
\par\end{centering}
\begin{centering}
\vspace{10pt}
\includegraphics[scale=0.5]{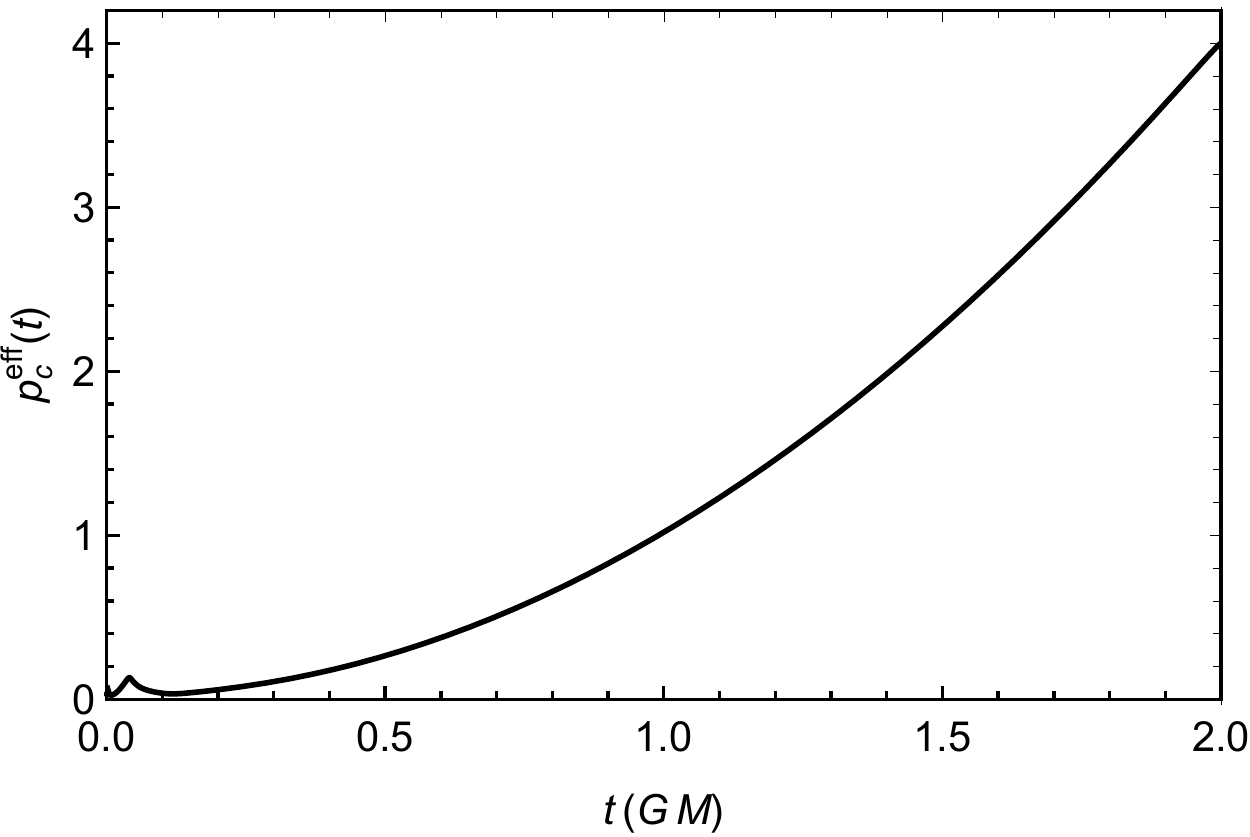}~~~~~~~~\includegraphics[scale=0.5]{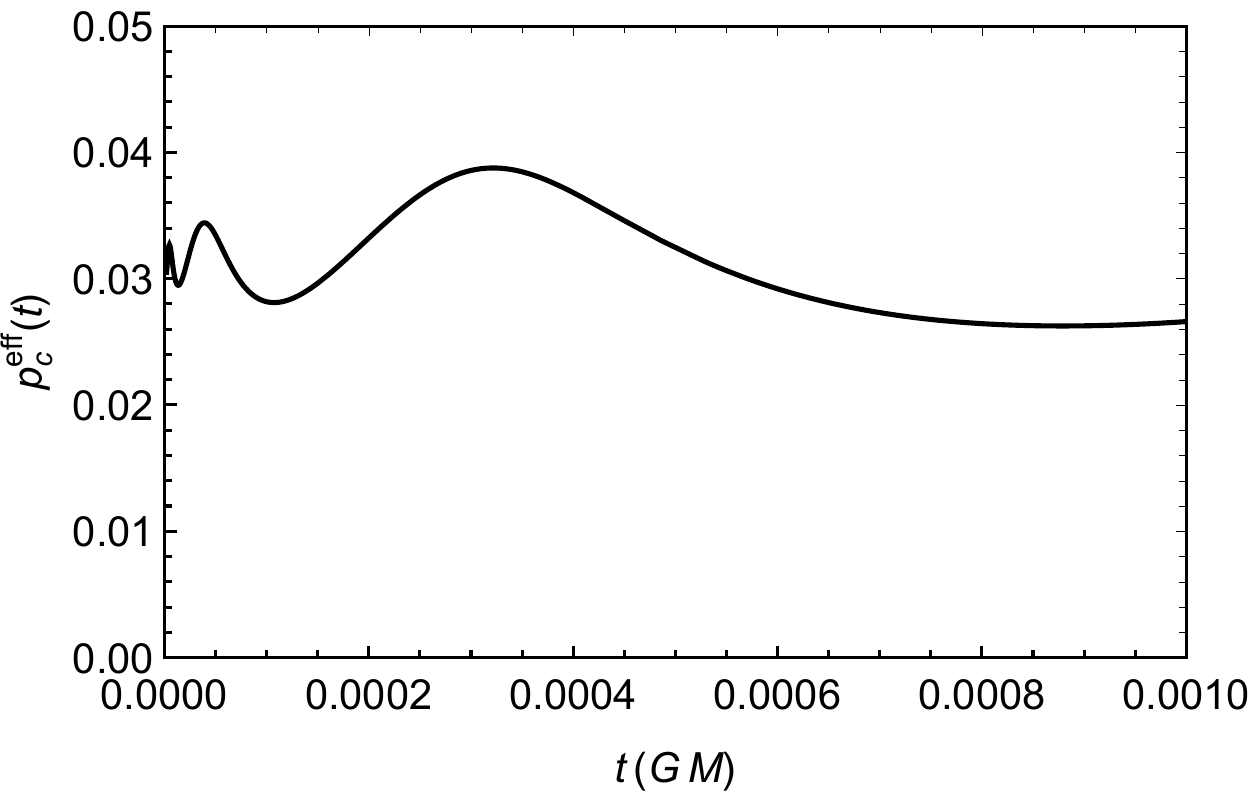}
\par\end{centering}
\caption{The behavior of the modified solutions in the $\bar{\mu}^{\prime}$ scheme
as a function of the Schwarzschild time $t$. The top left figure shows
$b(t)$ while top right figure shows the close-up of $b(t)$, close
to what used to be a singularity. The middle left figure depicts $c(t)$,
and the middle right one shows $p_{b}(t)$. Finally, the behavior of
$p_{c}$ and its close-up are depicted in bottom left and bottom right
figures, respectively. The figure is plotted using $\gamma=0.5,\,M=1,\,G=1,\,L_{0}=1$,
and $\Delta=0.1$. \label{fig:bar-prime-all-vars}}
\end{figure}

\subsection{$\bar{\mu}^{\prime}$ scheme\label{subsec:mubar-2-scheme}}


Here $\bar{\mu}_{b}^{\prime},\,\bar{\mu}_{c}^{\prime}$ have the following
dependence on the triad components,
\begin{align}
\bar{\mu}_{b}^{\prime}= & \sqrt{\frac{\Delta}{p_{c}}},\label{eq:mu-bar-b-prime}\\
\bar{\mu}_{c}^{\prime}= & \frac{\sqrt{p_{c}\Delta}}{p_{b}}.\label{eq:mu-bar-c-prime}
\end{align}
The equations of motion in this case are
\begin{align}
\frac{db}{dT}= & -\frac{1}{2}\gamma^{2}\frac{\bar{\mu}_{b}^{\prime}}{\sin\left(\bar{\mu}_{b}^{\prime}b\right)}-\frac{1}{2}\frac{\sin\left(\bar{\mu}_{b}^{\prime}b\right)}{\bar{\mu}_{b}^{\prime}}-\frac{p_{c}}{p_{b}}\left[\frac{\sin\left(\bar{\mu}_{c}^{\prime}c\right)}{\bar{\mu}_{c}^{\prime}}+c\cos\left(\bar{\mu}_{c}^{\prime}c\right)\right],\\
\frac{dp_{b}}{dT}= & \frac{1}{2}p_{b}\cos\left(\bar{\mu}_{b}^{\prime}b\right)\left[1-\gamma^{2}\frac{\bar{\mu}_{b}^{\prime2}}{\sin^{2}\left(\bar{\mu}_{b}^{\prime}b\right)}\right],\\
\frac{dc}{dT}= & \frac{p_{b}}{2p_{c}}\left[\gamma^{2}\frac{\bar{\mu}_{b}^{\prime}}{\sin\left(\bar{\mu}_{b}^{\prime}b\right)}\left[1-\frac{\bar{\mu}_{b}^{\prime}}{\sin\left(\bar{\mu}_{b}^{\prime}b\right)}b\cos\left(\bar{\mu}_{b}^{\prime}b\right)\right]-\frac{\sin\left(\bar{\mu}_{b}^{\prime}b\right)}{\bar{\mu}_{b}^{\prime}}\right]\nonumber \\
 & +\frac{bp_{b}\cos\left(\bar{\mu}_{b}^{\prime}b\right)}{2p_{c}}-\frac{\sin\left(\bar{\mu}_{c}^{\prime}c\right)}{\bar{\mu}_{c}^{\prime}}-c\cos\left(\bar{\mu}_{c}^{\prime}c\right),\\
\frac{dp_{c}}{dT}= & 2p_{c}\cos\left(\bar{\mu}_{c}^{\prime}c\right).
\end{align}
The behavior of some of these canonical variables as a function of
the Schwarzschild time $t$ is now quite different from the previous
two schemes. Figure \ref{fig:bar-prime-all-vars} shows the behavior
of $b,\,p_{c}$ and their close-ups near what used to be a singularity
as well as the behavior of $c,\,p_{b}$. Particularly, one notices
that $p_{c}$ behaves differently while still remaining nonzero in
the interior, hence leading singularity resolution once again. Both
$b$ and $p_{c}$ show some sort of damped oscillatory behavior close
to the classical singularity which contributes to a more volatile behavior
of the Raychaudhuri equation. 

\begin{figure}
\begin{centering}
\includegraphics[scale=0.5]{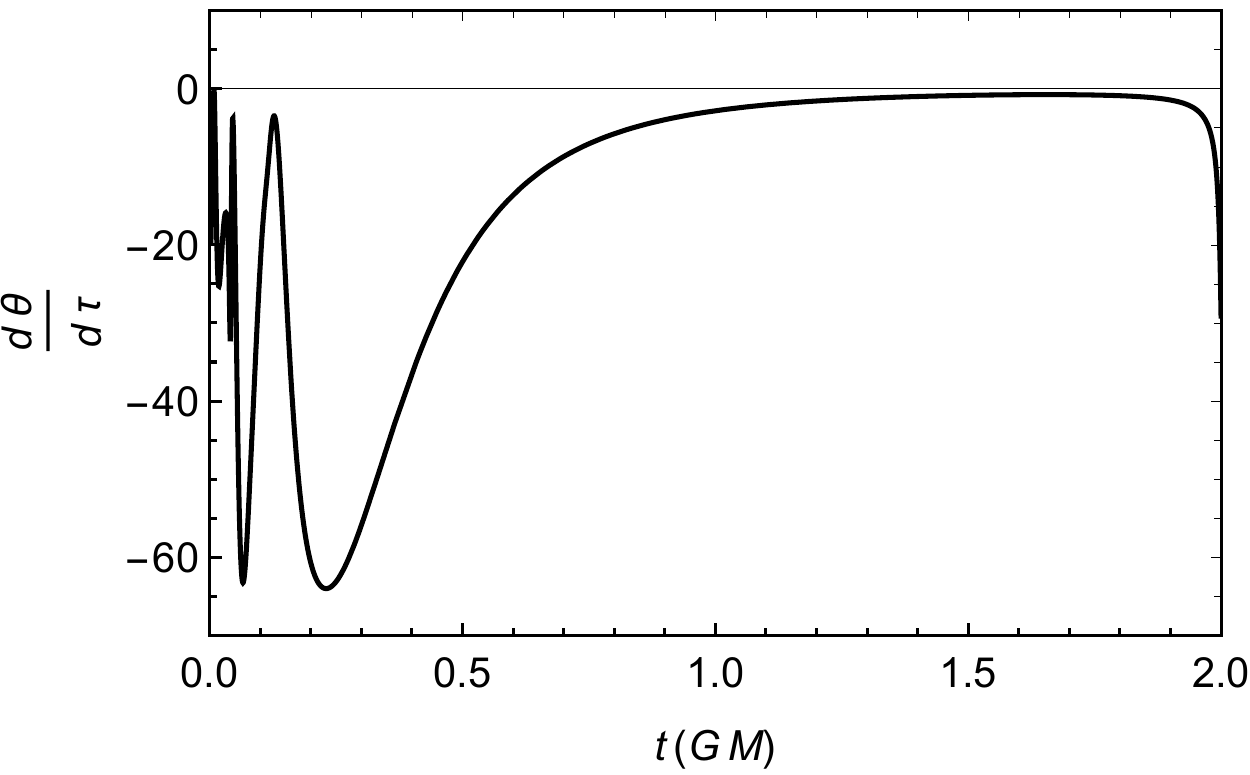}~~~~~~~~\includegraphics[scale=0.5]{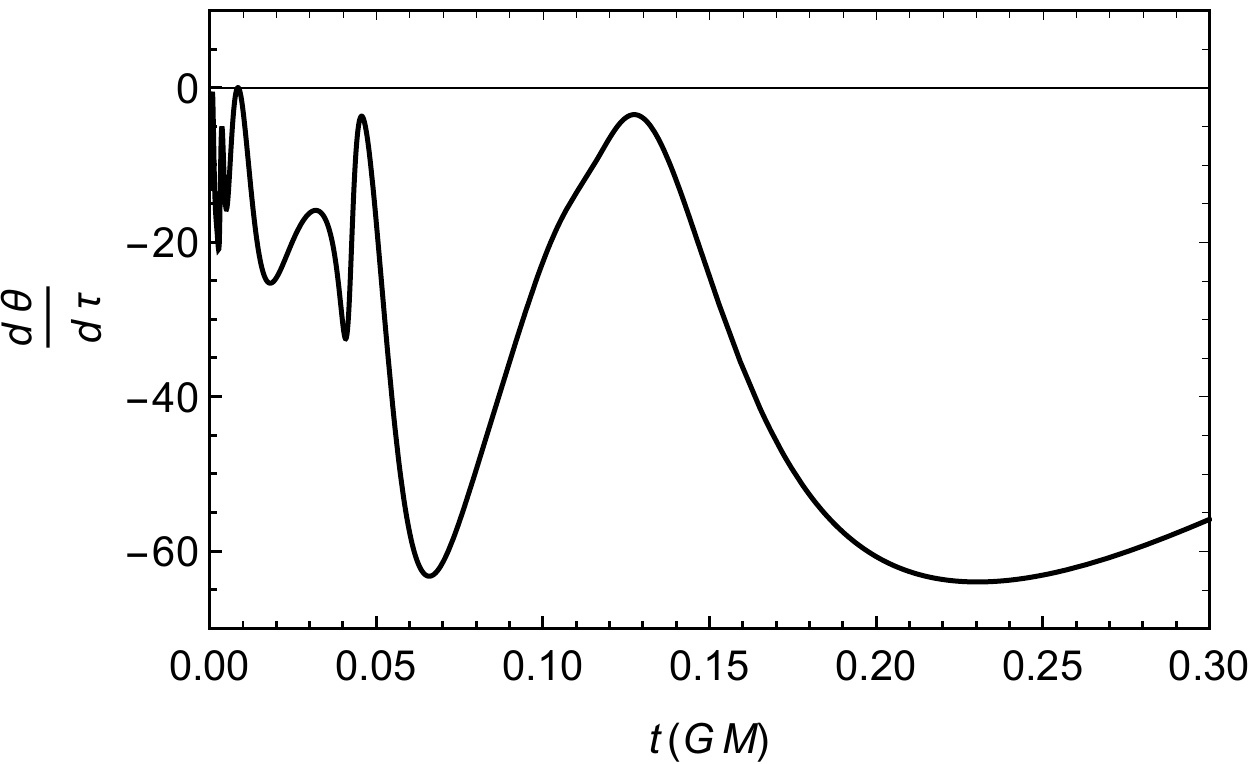}
\par\end{centering}
\begin{centering}
\vspace{10pt}
\includegraphics[scale=0.5]{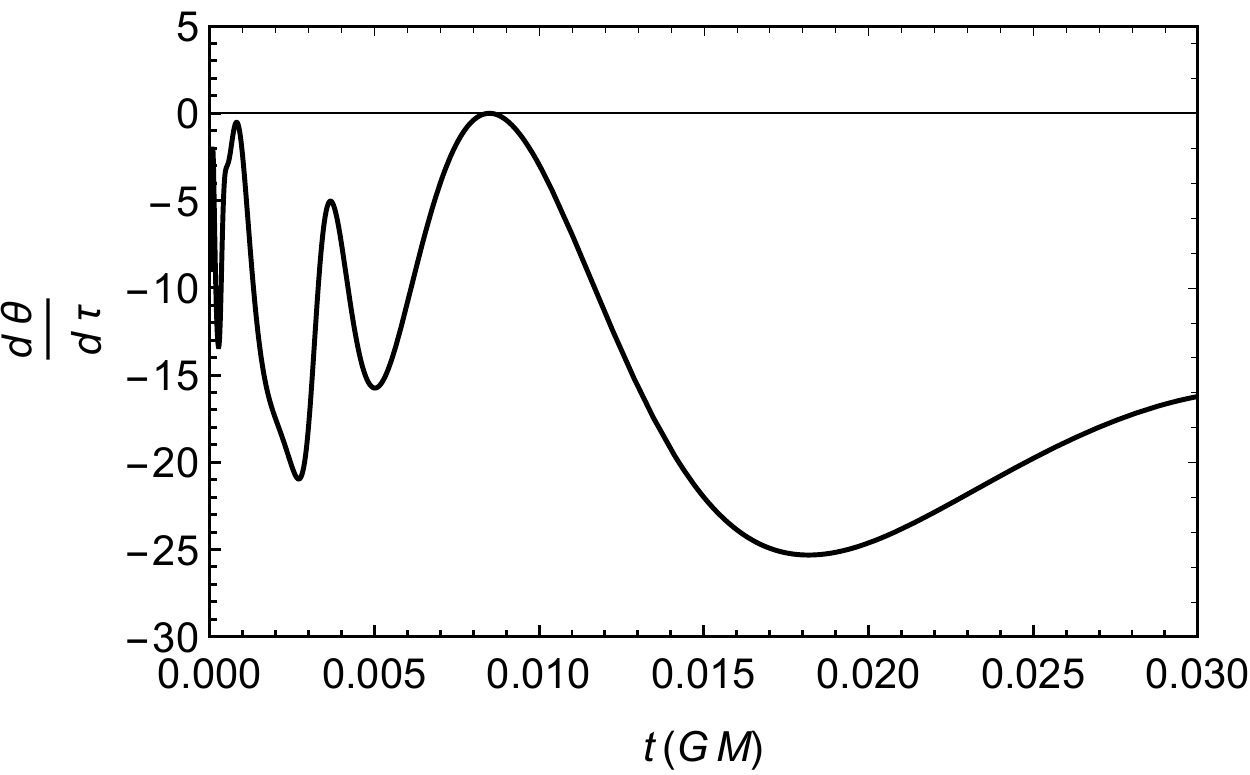}~~~~~~~~\includegraphics[scale=0.5]{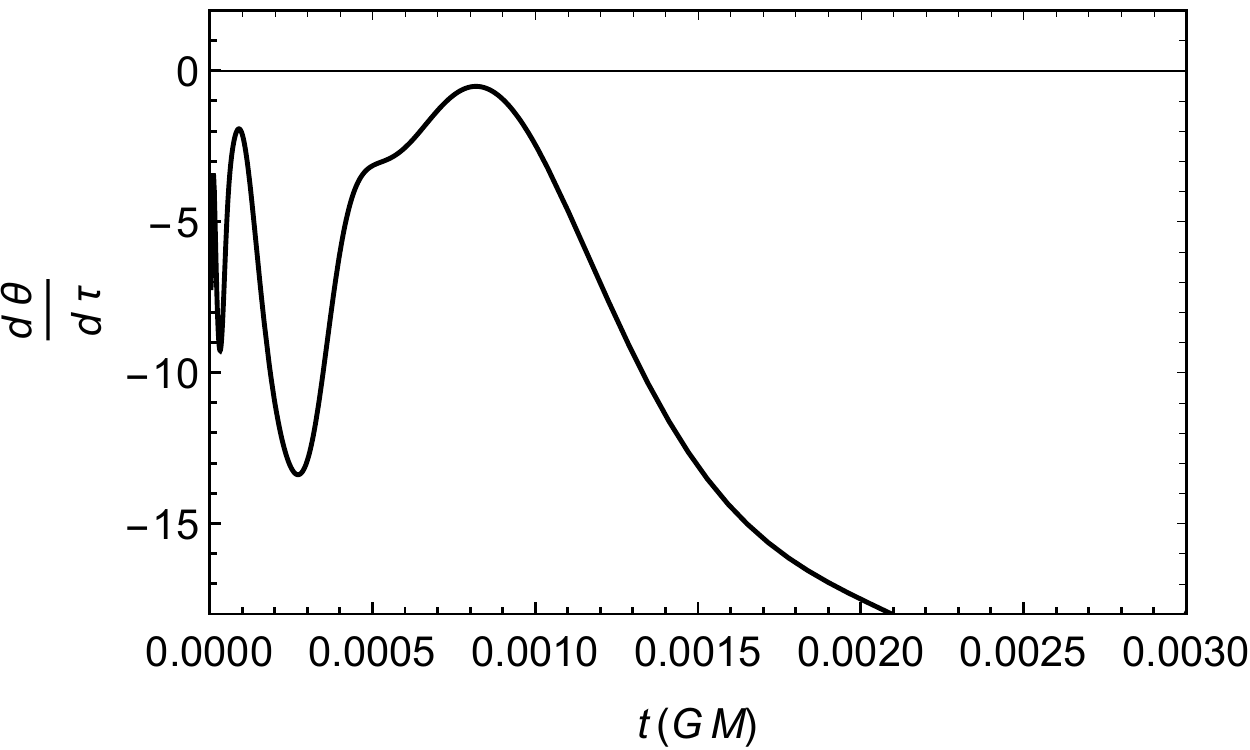}
\par\end{centering}
\begin{centering}
\vspace{10pt}
\includegraphics[scale=0.5]{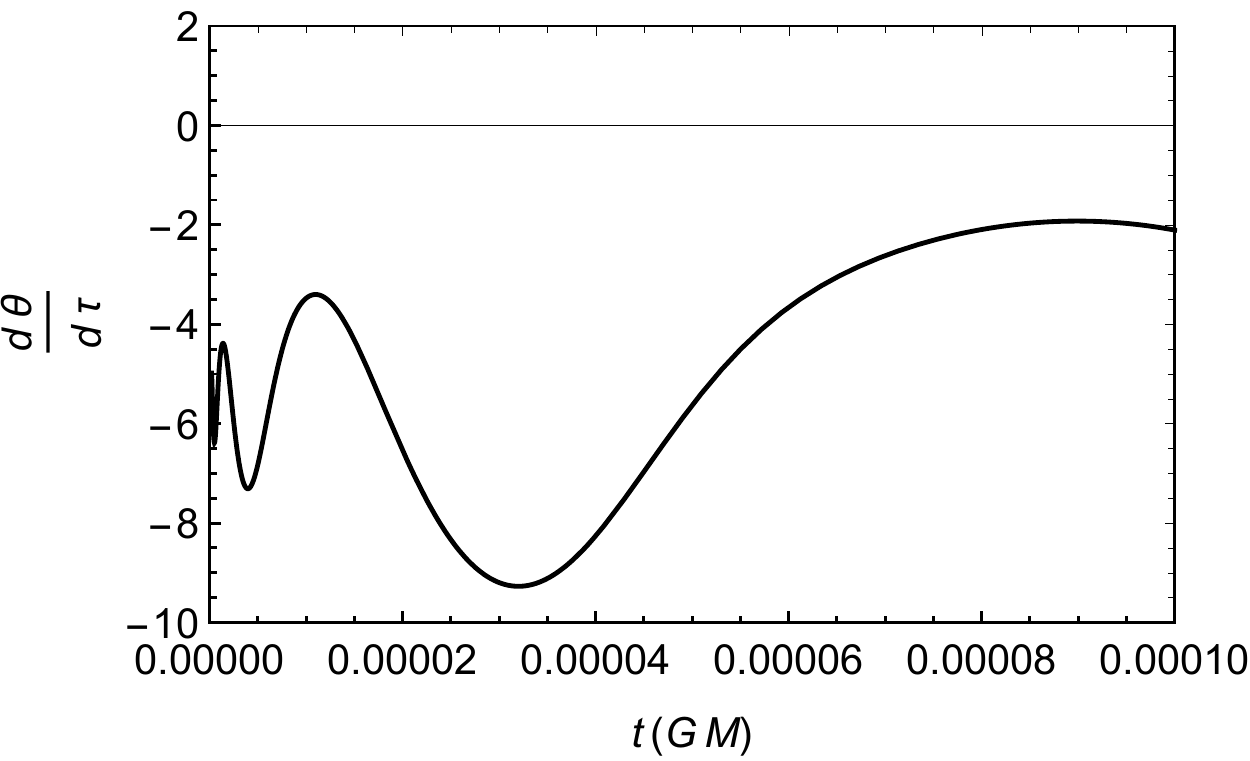}~~~~~~~~\includegraphics[scale=0.5]{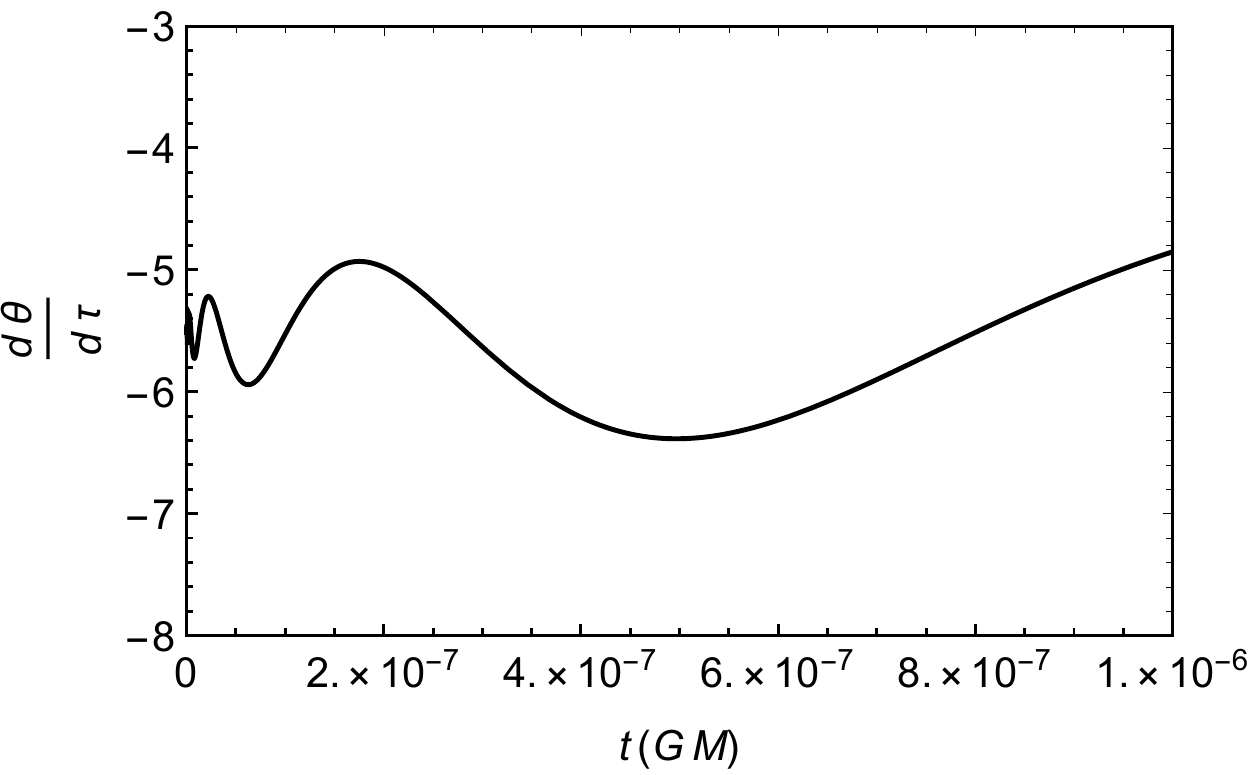}
\par\end{centering}
\caption{Raychaudhuri equation in the $\bar{\mu}^{\prime}$ scheme. The top left
figure shows the behavior over the whole $0\protect\leq t\protect\leq2GM$
range. Other plots show various close-ups of that plot over smaller
ranges of $t$. The figure is plotted using $\gamma=0.5,\,M=1,\,G=1,\,L_{0}=1$,
and $\Delta=0.1$. \label{fig:RE-bar-prime}}
\end{figure}

The full Raychaudhuri equation in this case takes the following form
\begin{align}
\frac{d\theta}{d\tau}= & \frac{1}{\gamma^{2}p_{c}}\frac{\sin^{2}\left(b\bar{\mu}_{b}^{\prime}\right)}{\bar{\mu}_{b}^{\prime2}}\left[\cos\left(\bar{\mu}_{b}^{\prime}b\right)\cos\left(\bar{\mu}_{c}^{\prime}c\right)-\frac{\cos^{2}\left(\bar{\mu}_{b}^{\prime}b\right)}{4}-3\cos^{2}\left(\bar{\mu}_{c}^{\prime}c\right)\right]\nonumber \\
 & +\frac{\cos\left(\bar{\mu}_{b}^{\prime}b\right)}{p_{c}}\left[\frac{\cos\left(\bar{\mu}_{b}^{\prime}b\right)}{2}-\cos\left(\bar{\mu}_{c}^{\prime}c\right)-\frac{\gamma^{2}}{4}\cos\left(\bar{\mu}_{b}^{\prime}b\right)\frac{\bar{\mu}_{b}^{\prime2}}{\sin^{2}\left(\bar{\mu}_{b}^{\prime}b\right)}\right].
\end{align}
This again looks identical to (\ref{eq:dtheta-dtau-mu0})
and (\ref{eq:dtheta-dtau-mu-bar-1}), except for the different forms of $\bar{\mu}^{\prime}$
scales compared to previous cases. Before considering the full nonperturbative
expression of the above equation, we can check that up to first order
in $\Delta$, one obtains
\begin{equation}
\frac{d\theta}{d\tau}\approx-\frac{1}{2p_{c}}\left(1+\frac{9b^{2}}{2\gamma^{2}}+\frac{\gamma^{2}}{2b^{2}}\right)+\frac{\Delta}{6\gamma^{2}}\left[\frac{1}{p_{c}^{2}}\left(3b^{4}+\gamma^{4}\right)+\frac{3c^{2}}{p_{b}^{2}}\left(5b^{2}+\gamma^{2}\right)\right].
\end{equation}
Although this perturbative form of the Raychaudhuri equation is a bit different from previous cases, nevertheless it exhibits the property that the quantum corrections are all positive and hence once again contribute to defocusing of the geodesics. 

The full nonperturbative Raychaudhuri equation and its close-ups in
this case are plotted in Fig. \ref{fig:RE-bar-prime}. It is seen
that in this scheme, the Raychaudhuri equation exhibits a more volatile
behavior and has various bumps particularly when we get closer to
where the singularity used to be. Very close to the classical singularity,
its form resembles those of $b$ and $p_{c}$, behaving like a damped
oscillation.

Two particular features are worth noting in this scheme. First,
as we also saw in previous schemes, quantum corrections kick in close
to the singularity and dominate the evolution such that the infinite
focusing is remedied, hence signaling the resolution of the singularity.
Second, this scheme exhibits a nonvanishing value for $\frac{d\theta}{d\tau}$
at or very close to the singularity. In Fig. \ref{fig:RE-bar-prime}
with the particular choice of numerical values of $\gamma,\,M,\,G,\,L_{0}$
and $\Delta$, the value of $\frac{d\theta}{d\tau}$ for $t\to0$
is approximately $-5.5$. Hence, although a nonvanishing focusing
is not achieved in this case at where the singularity used to be,
nevertheless, there exists a relatively small focusing.

\section{Discussion and outlook}

In this paper, we have shown that the LQG corrections to the interior of Schwarzschild black hole induce additional terms in the Raychaudhuri equation. Importantly, these terms are repulsive (positive) near the classical singularity of the black hole. This is in contrast to the attractive (negative) terms on the right-hand side of the classical Raychaudhuri equation. So, while the former implies the convergence of geodesics, as our explicit computation and related plots show, the quantum generated terms that we estimated are sufficient to negate this convergence. 
This in turn implies that the primary condition for the
Hawking-Penrose singularity theorems to hold true is violated, and the theorems themselves cease to hold. Consequently, geodesics are no longer incomplete, and the classical singularity is resolved. We emphasize that this result is true only for the spherically symmetric Schwarzschild black hole, but we expect it to continue to hold for realistic astrophysical black holes as well, with little or no symmetries. After all, it is the more symmetric solutions which are more likely to demonstrate singularities. While repeating our calculations for the most general black hole singularity may prove technically challenging, we do hope to extend our results to other black hole spacetimes, such as Kerr or Reissner-Nordstr\"om. Furthermore, our approach should shed light on cosmological singularities as well. We hope to report on these elsewhere.

\acknowledgments{
This work was supported by the 
Natural Sciences and Engineering Research Council (NSERC) of Canada. 
}

\appendix

\bibliographystyle{JHEP}
\bibliography{mainbib}

\end{document}